\documentclass{aa}  
\usepackage{CJKutf8}
\usepackage{graphicx}
\usepackage[breaklinks,colorlinks,citecolor=blue,linkcolor=blue]{hyperref}
\usepackage{txfonts}
\usepackage{lscape}
\usepackage{epstopdf}
\usepackage{CJK}
\usepackage{float}
\usepackage[absolute]{textpos}
\usepackage{amsmath}
\usepackage{mathrsfs}

\usepackage{arydshln}

\begin{document}

   \title{Sulfur oxides tracing streamers and shocks at low mass
   protostellar disk-envelope interfaces}


   \author{X.-C. Liu \begin{CJK}{UTF8}{gbsn}(刘训川)\end{CJK}
          \inst{1,3} 
          \and
          E. F. van Dishoeck
          \inst{1}
          \and
          M. R. Hogerheijde \inst{1,2}
          \and
          M. L. van Gelder
          \inst{1}
          \and
          Y. Chen\inst{1}
          \and
          T. Liu\inst{3}
          \and
          M. van't Hoff\inst{4}
          \and
          M. N. Drozdovskaya\inst{5}
          \and
          E. Artur de la Villarmois\inst{6}
          \and
          X.-F. Mai\inst{7,3}
          \and 
          Ł. Tychoniec\inst{1}
          }

   \institute{Leiden Observatory, Leiden University, P.O. Box 9513, 2300RA
Leiden, The Netherlands\\
              \email{liuxunchuan001@gmail.com}
              \and 
              Anton Pannekoek Institute for Astronomy, University of Amtserdam, the Netherlands.
              \and              
              Shanghai Astronomical Observatory, Chinese Academy of Sciences, 80 Nandan Road, Shanghai 200030, PR China
              \and 
              University of Michigan, 323 West Hall, 1085 South University Ave., Ann Arbor, MI 48109, USA
              \and
               Physikalish-Meteorologisches Observatorium Davos und Weltstrahlungszentrum (PMOD/WRC), Dorfstrasse 33, 7260 Davos Dorf,
               Switzerland
              \and
              European Southern Observatory, Alonso de Córdova 3107, Casilla 19, Vitacura, Santiago, Chile
              \and
              Department of Physics, PO Box 64, 00014, University of Helsinki, Finland
             }

   \date{Received xxx xx}

 
  \abstract
{Accretion shocks are thought to play a crucial role in the early stages of star and planet formation, but their direct observational evidence remains elusive, particularly regarding the molecular tracers of these processes. In this work, we searched for features of accretion shocks by observing the emission of SO and SO$_2$ using ALMA in Band 6 towards nearby Class I protostars. We analyze the SO and SO$_2$ emission from Oph IRS 63, DK Cha, and L1527, which have different disk inclination angles, ranging from nearly face-on to edge-on. SO emission is found to be concentrated in rings at the centrifugal barriers of the infalling envelopes. These rings are projected onto the plane of the sky as ellipses or parallel slabs, depending on the inclination angles. Spiral-like streamers with SO emission are also common, with warm ($T_{\rm ex} > 50$ K) and even hot ($T_{\rm ex} \gtrsim 100$ K) spots or segments of SO$_2$ observed near the centrifugal barriers.  Inspired by these findings, we present a model that consistently explains the accretion shock traced by SO and SO$_2$, where the shock occurs primarily in two regions: (1) the centrifugal barriers, and (2) the surface of the disk-like inner envelope outside the centrifugal barrier. The outer envelope gains angular momentum through outflows, causing it to fall onto the midplane at or outside the centrifugal barrier, leading to a disk-like inner envelope that is pressure-confined by the accretion shock and moves in a rotating-and-infalling motion. We classify the streamers into two types—those in the midplane and those off the midplane. These streamers interact with the inner envelopes in different ways, resulting in different patterns of shocked regions. We suggest that the shock-related chemistry at the surfaces of the disk and the disk-like inner envelope warrants further special attention.
}


   \maketitle
%

\section{Introduction}\label{sec_intro}
Protostars, the progenitors of the central stars in new solar systems, arise from the gravitational collapse of gas and dust in their natal clouds \citep[e.g.,][]{1969MNRAS.145..271L,1977ApJ...214..488S}. Due to angular momentum, mass accretion leads to the development of a rotating, infalling envelope and a rotation-dominated disk \citep[e.g.,][]{1976ApJ...210..377U,1987ApJ...312..788A}. 
One of the major questions is to what extent the chemical composition is preserved from cloud to disk (referred to as ‘inheritance’), or whether it is modified along the way (referred to as ‘modification’) \citep[e.g.,][]{1999ApJ...519..705A,2009A&A...495..881V,2010A&A...519A..28V,2014MNRAS.445..913D,2023ApJ...956..120H}. The strong similarity between interstellar and cometary ices suggests that at least some inheritance from the interstellar medium (ISM) occurs \citep{2000A&A...353.1101B,2004come.book..391B,2011ARA&A..49..471M,2019MNRAS.490...50D}. 
At the same time, there is growing evidence that planetesimal formation begins early, even during the embedded phase of star formation \citep{2010A&A...510A..72F,2018A&A...618L...3M,2020A&A...640A..19T}. This process may preserve the original chemical composition by trapping molecules in large icy bodies that no longer participate in the chemistry on small dust grains \citep[e.g.,][]{2016ApJ...823L..41M,2016MNRAS.462S..99T,2019A&A...621A..75E}. A key uncertainty in this process, however, is whether most of the material undergoes a strong accretion shock before or as it enters the disk, which could potentially ‘reset’ the chemical composition.

The infall of material in the envelope can cause a shock when it collides with/around the disk, raising the temperatures of gas and dust to levels much higher than those produced by stellar photon heating alone \citep[e.g.,][]{1983ApJ...264..485D,1994ApJ...428..170N}. However, to date, there is limited definitive observational evidence confirming the existence of strong accretion shocks. Typical shock tracers in the optical and infrared wavelengths (e.g., \mbox{[O {I}] 6300~\AA}, \mbox{[S {II}] 6731~\AA}, \mbox{[S I] 25 $\mu$m}, \mbox{[O I] 63 $\mu$m}, \citealt{2011A&A...527A..13P,2019ApJ...870...76B,2015ApJ...801..121N,2016A&A...594A..59R}) are more sensitive to high-velocity shocks (10 km s$^{-1}$ or higher). These tracers are often contaminated by outflows and/or suffer from high extinction due to the surrounding envelope. Furthermore, outflows and jets are very common in protostars, even during the very early first hydrostatic core (FHSC) phase \citep{2015A&A...577L...2G,2020A&A...633A.126B}, making it unclear whether or not they prevent the direct accretion of material onto the inner regions of the disks.

\begin{table*}[!thb]
\centering
\caption{The targeted lines$^{(a)}$ of SO and SO$_2$.\label{tab_transitions}}
{
\centering
\begin{tabular}{ccccclcc}
\hline\hline 
Mol. & Trans.$^{(b)}$ & Freq. & $S_{ij}\mu^2$ &  $E_{\rm up}$ & ALMA Projects and targets$^{(c)}$ & Peak S/N$^{(d)}$ &  Total flux$^{(f)}$\\
        &   $N_J$ ($J_{K_a,K_c}$)   & GHz       & ${\rm D}^2$         &  K && & mJy km s$^{-1}$\\
\hline
SO     & $5_5 - 4_4$   & 215.22065 & 11.5  & 44.1  & 2023.1.00592.S (L1527) & 20 &  1030$\pm$16 \\
       & $1_2-2_1$     & 236.45229 & 0.03  & 15.8  & 2022.1.01411.S (IRS 63, DK Cha) & -- & --\\
       & $5_6-4_5$     & 251.82577 & 11.6  & 50.6  & 2022.1.01411.S (IRS 63, DK Cha) & 27, 26 &  6200$\pm$50, 6300$\pm$80\\ 
\hline
SO$_2$ & $16_{3,13}-16_{2,14}$  & 214.68938 & 28.4 & 148  & 2023.1.00592.S (L1527) & 3 & 87$\pm$10\\
       & $16_{1,15}-15_{2,14}$  & 236.21668 & 16.1 & 131  & 2022.1.01411.S (IRS 63, DK Cha) & 3, 3 & 105$\pm$11, 480$\pm$30  \\ 
       & $_{\ }12_{3,9}-12_{2,10}$   & 237.06887 & 18.4 & 93.9 & 2022.1.01411.S (IRS 63, DK Cha) & 3.5, 4 & 290$\pm$25, 590$\pm$35 \\ 
       & $15_{2,14}-15_{1,15}$  & 248.05740 & 14.1 & 119  & 2022.1.01411.S (IRS 63, DK Cha) & 2.5, 3 &42$\pm$6, 510$\pm$32 \\ 
       & $13_{1,13}-12_{0,12}$  & 251.19967 & 25.7 & 82.2 & 2022.1.01411.S (IRS 63, DK Cha) & 7, 7 & 640$\pm$25, 1020$\pm$30\\ 
       & $8_{3,5}-8_{2,6}$      & 251.21058 & 11.0 & 55.2 & 2022.1.01411.S (IRS 63, DK Cha) & 5, 4 & 600$\pm$23, 750$\pm$30\\ 
\hline
\end{tabular}
}\\
{\small 
\begin{flushleft}
$^{(a)}$ The transition parameters are quoted from the JPL \citep{1998JQSRT..60..883P}
and CDMS \citep{2016JMoSp.327...95E} database through
Splatalogue \protect\footnotemark.
The columns from the left to right list the 
names of molecules (Mol.), transition quantum numbers (Trans.), 
rest frequency (Freq.), line strengths ($S_{ij}\mu^2$),
upper-level energies ($E_{\rm up}$) and some notes (Note).
The unit of $S_{ij}\mu^2$ is Debye$^2$ (denoted as ${\rm D}^2$).
Note that transitions of molecules that have been covered by the observations 
but were not used in this work are not listed.
$^{(b)}$ For both the upper and lower levels,
the transition labels for SO and SO$_2$ are $N_J$ and $J_{K_a,K_c}$,
respectively.
$^{(c)}$ Only the targets relevant to this work are listed.
IRS 63 represents Oph IRS 63.
$^{(d)}$ The signal-to-noise ratios (S/Ns) are measured from spectra extracted at the emission peaks for the corresponding targets listed in the sixth column, under the native angular and spectral resolution, when cleaned with a \textit{robust} parameter of 0.5. The S/Ns can be improved through smoothing. The symbol `--' indicates non-detection.
$^{(f)}$ The total fluxes and their uncertainties for the corresponding targets listed in the sixth column, obtained by applying Gaussian fitting to the integrated spectrum over the whole emission region.
\end{flushleft}
}
\end{table*}

Sulfur oxides (SO and SO$_2$) are commonly used as molecular tracers of shocks \citep[e.g.,][]{1997ApJ...487L..93B,2007ApJ...659..499L,2022A&A...658A.104G,2022ApJS..263...13L,2024A&A...682A..78V}. At typical shock speeds (ranging from several to \mbox{$\sim10$ km s$^{-1}$}) for impacting gas with densities greater than \mbox{$10^7$ cm$^{-3}$}, the molecules largely survive, but the gas temperatures just behind the shock can reach thousands of Kelvin, which can drive most of the oxygen into OH and H$_2$O \citep[e.g.,][]{2009A&A...495..881V}. This, in turn, leads to enhanced abundances of molecules like SO and SO$_2$ \citep{1994ApJ...428..170N,2017ApJ...839...47M,2021A&A...653A.159V}. SO and SO$_2$ produced by outflows are expected to exist on scales much larger than the disk size \citep{2005A&A...437..149W,2020ApJ...896...37F,2021ApJ...912..148L}, and it should be possible to distinguish these from molecular enhancements induced by accretion shocks. 
Sulfur-bearing molecules appear to be particularly promising as tracers of energetic processes at the disk-envelope interface \citep[e.g.,][]{2014Natur.507...78S,2021A&A...646A..72H,2022A&A...658A.104G,2024A&A...689L...7D}.

Advanced interferometers, especially the Atacama Large Millimeter/submillimeter Array (ALMA), offer unprecedented opportunities to uncover systems of envelopes, disks, and outflows with exceptional spatial resolution, i.e., tens of AU \citep{2012Natur.492...83T,2015ApJ...808L...3A,2016Natur.540..406B,2019FrASS...6....3H,2023ApJ...951....8O}.
This capability enables the detection of shock-related sulfur-bearing molecules around disks.
\citet{2014Natur.507...78S,2017MNRAS.467L..76S} detected a parallel slab of SO toward L1527, an edge-on low-mass system, which shows a strong decreasing trend in the position-velocity map and interpreted it as shock tracers occurring around and inside the centrifugal barrier. At the inner radius of the SO ring of AB Aurigae, a Herbig Ae star with a nearly face-on protoplanetary disk, was found to trace gas located, in part, beyond the dust ring \citep{2024A&A...689L...7D}. The merging zone of the spiral structures and the disk of AB Aurigae may trigger gravitational instability, potentially leading to planet formation \citep{2025ApJ...981L..30S}. In addition to SO, strong SO$_2$ emission has also been detected on disk scales or in the infalling streamers of young protostars \citep{2022A&A...667A..12V,2022A&A...658A.104G,2024A&A...682A..78V}. These observations suggest that SO and SO$_2$ may be strongly linked to accretion shocks \citep{2019A&A...626A..71A,2022A&A...667A..20A}. However, other studies have found that SO can also trace the volatile reservoir during the epoch of planet and comet formation \citep{2023A&A...669A..53B,2023NatAs...7..684K,2024AJ....167..165B}. The detailed origin of SO and its relation to SO$_2$ remain unclear.
Furthermore, a unified model that explains the accretion shock at the interface between the envelope and the disk is still lacking.

Overall, high-resolution observations of the emission lines from both SO and SO$_2$ are crucial for assessing whether their emission can be attributed to accretion shocks and for investigating the details of how and where this process occurs. To address these issues, we carried out ALMA observations of SO and SO$_2$ towards several disk-hosting protostars with varying inclination angles. We examined the low-mass data for signatures of accretion shocks and sought to explain the observational results for shock tracers (especially SO and SO$_2$) in this study, as well as those in the literature, within the framework of a consistent model of disk-envelope interface. This work is structured as follows: The ALMA observations are described in Sect. \ref{sec_obs}; the observational results and relevant analyses are presented in Sect. \ref{sec_obsresults}; these data trigger the development of a new semi-analytic model to explain the observations, which is presented in Sect. \ref{sec_model}; and the discussion and brief summary are presented in Sects. \ref{sec_discussion} and \ref{sec_summary}, respectively.


\section{Observations and data}\label{sec_obs}
\subsection{Observations}
\subsubsection{ALMA observational set up}\label{sec_obssetup}
The ALMA data analyzed in this paper are primarily from programs 2022.1.01411.S and 2023.1.00592.S (PI: M. L. van Gelder), which aim to search for signatures of accretion shocks in protostars.
This study examines three nearby protostars (see Sect. \ref{sec_sample} for details) from the two projects.
Both programs covered several transitions of SO and SO$_2$ in ALMA Band 6, but with different frequency setups
(Table \ref{tab_transitions}). The rest frequencies ($f_{\rm rest}$) of the lines targeted by program 2022.1.01411.S are within 233–252 GHz, while program 2023.1.00592.S targets lines within a slightly lower range of 215–232 GHz.
The channel width is 0.141 MHz, corresponding to a velocity resolution of 0.17 km s$^{-1}$ at 250 GHz and 0.19 km s$^{-1}$ at 220 GHz. Each spectral window (SPW) has a bandwidth of 58.59 MHz (corresponding to $\sim$70 km s$^{-1}$ at 250 GHz). Note that the two SO$_2$ lines ($8_{3,5}-8_{2,6}$ and $13_{1,13}-12_{0,12}$) are separated by only $\sim$11 MHz ($\sim$13 km s$^{-1}$) and are therefore covered within the same SPW.


Two 12-m array configurations, C6 and C3, were employed for both programs. The maximum recoverable scales (MRS) are $\sim$2.5\arcsec~and $\sim$7\arcsec~for the C6 and C3 configurations, respectively. \footnotetext{\url{https://splatalogue.online/}} The data were pipeline-calibrated and imaged using the Common Astronomy Software Applications\footnote{\url{https://casa.nrao.edu/}} \citep[CASA;][]{2007ASPC..376..127M} version 6.4.1.12. To resolve disk-scale structures while maintaining sensitivity, we combined the 12-meter array data using Briggs weighting with a robust parameter of 0.5 — the default in CASA's tclean task, as it offers an optimal balance between resolution and sensitivity.
For L1527, to enhance the signal-to-noise ratio (S/N) of its large-scale SO emission associated with the outflow (Sect. \ref{sec_L1527}), we additionally applied {\it natural} weighting, which resulted in a lower noise level but poorer resolution. In CASA's \texttt{tclean}, \textit{Briggs} weighting with a \textit{robust} parameter of 2 is effectively equivalent to natural weighting. 
The multiscale deconvolver was adopted with \textit{auto-multithresh} masking, using \texttt{sidelobethreshold} of 2.0 and \texttt{noisethreshold} of 4.25. The Full Width at Half Maximum (FWHM) of the synthesized beams depends on the declination of the targets and the weighting schemes used during the cleaning process. It has values of $\sim 0.15\arcsec$–$0.2\arcsec$ for \mbox{{\it robust} = 0.5} and $\sim 0.35\arcsec$ for \mbox{{\it robust} = 2}. The 1-$\sigma$ noise level at the native spectral resolution (0.141 MHz), when cleaned with a {\it robust} parameter of 0.5, is approximately 2.5 mJy beam$^{-1}$ and 2.0 mJy beam$^{-1}$ for programs 2022.1.01411.S and 2023.1.00592, respectively. A flux calibration uncertainty of 5\% was assumed \citep{2020AJ....160..270F}. The main data used for this work are the ALMA observations of three young protostars (Oph IRS 63, DK Cha, and L1527) with different inclination angles (see details in Sect. \ref{sec_sample}). Note that an angular resolution of 0.15\arcsec~corresponds to a spatial resolution of approximately 23~au, 27~au, and 22~au for IRS 63, DK Cha, and L1527, respectively.

\begin{figure*}
\centering
\includegraphics[width=0.995\linewidth]{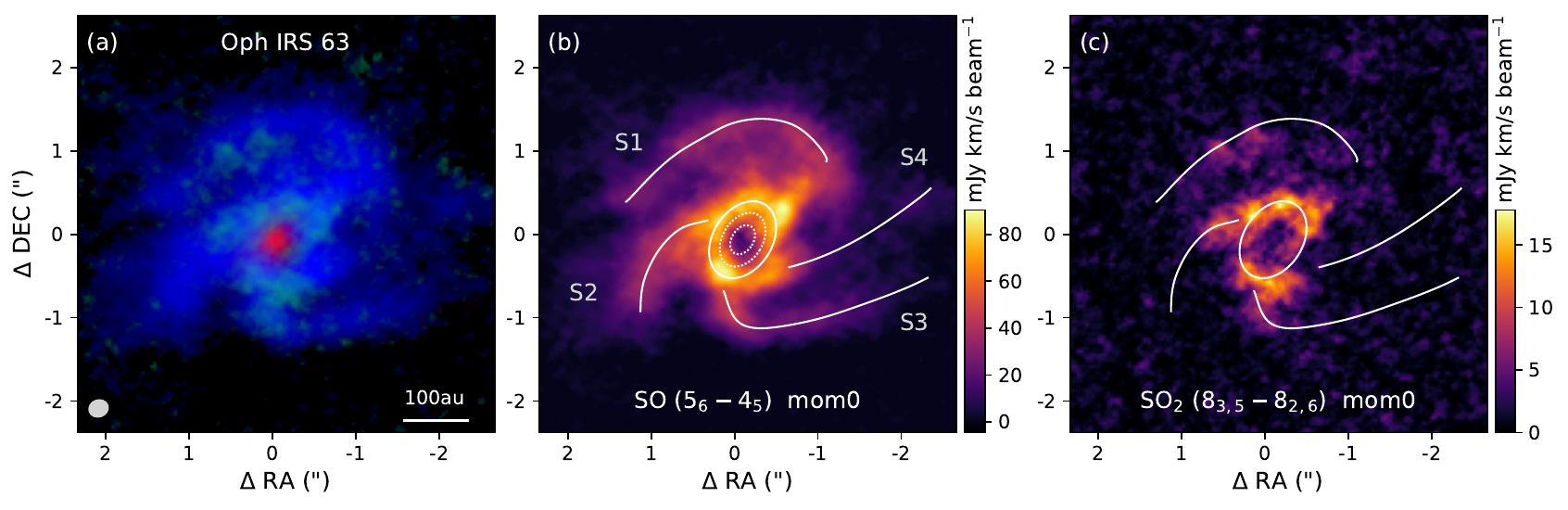}
\caption{(a) Three-color image of Oph IRS 63, where red, blue, and green represent emissions from the 1.3 mm continuum, SO ($5_6-4_5$) \mbox{(moment 0)}, and \mbox{SO$_2$ ($8_{3,5}-8_{2,6}$)} \mbox{(moment 0)}, respectively. The synthesized beam is shown in the lower left corner. (b) Moment 0 map of SO ($5_6-4_5$) toward Oph IRS 63. The two dashed white ellipses denote the two dust rings of the protoplanetary disk revealed by \citet{2020Natur.586..228S}. The solid white ellipse represents the SO ring. Four spiral structures (S1 to S4) are shown as white curves. (c) Moment 0 map of SO$_2$ ($8_{3,5}-8_{2,6}$) toward Oph IRS 63. The white ellipse and curves have the same meanings as in panel (b). Note that the images shown here represent only a portion of the field of view of the observation and are centered at RA (J2000) = 16$^{\rm h}$31$^{\rm m}$35.65$^{\rm s}$, Dec (J2000) = $-24^{\rm d}$01$^{\rm m}$30.1$^{\rm s}$.
\label{fig_irs63_color_maps}
}
\end{figure*}

\begin{figure}[!t]
\centering
\includegraphics[width=0.95\linewidth]{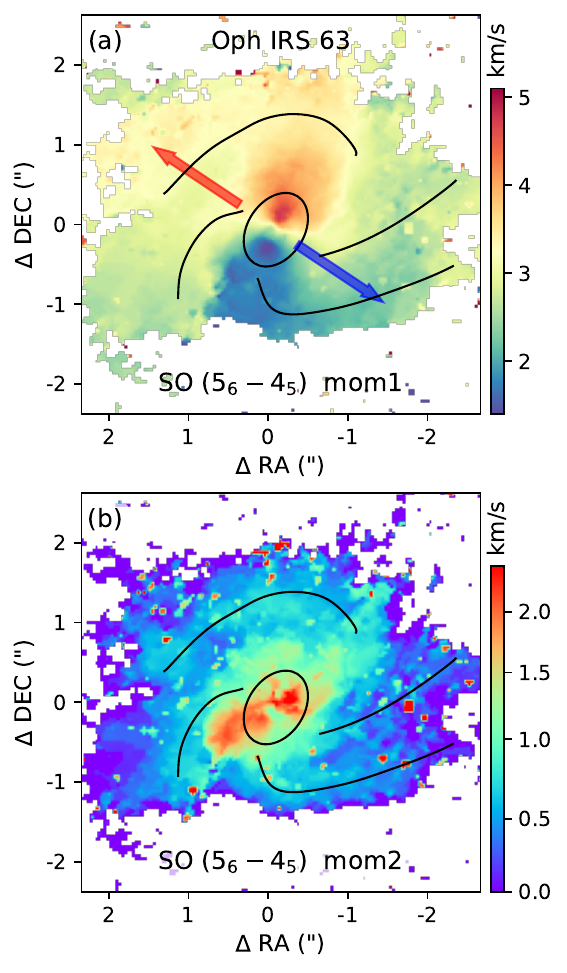}
\caption{Panels (a) and (b) show the moment 1 and moment 2 maps of SO ($5_6-4_5$) toward Oph IRS 63, respectively. In panel (a), the directions of the red and blue outflows \citep{2023ApJ...958...98F} are indicated by the arrows in the corresponding colors. In panel (b), the original moment 2 map has been multiplied by a factor of $\sqrt{8\ln(2)}$ and should be interpreted as the line width. The black ellipse and curves denote the SO rings and spiral structures (Fig. \ref{fig_irs63_color_maps}).
\label{fig_irs63_SO_mom12}
}
\end{figure}

\begin{figure}[!t]
\centering
\includegraphics[width=0.99\linewidth]{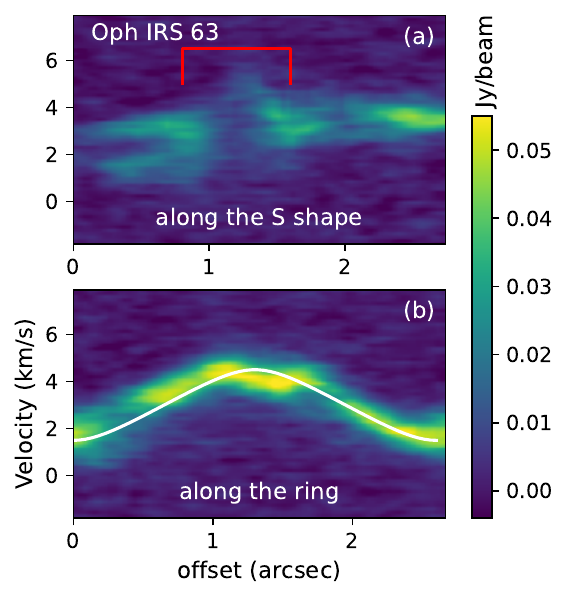}
\caption{
(a) PV map of SO along the S-shaped structure observed in the moment maps shown in Fig. \ref{fig_irs63_SO_mom12} (see Fig. \ref{fig_irs63_Smark} for the implicit marking of the S-shape).
The red line marks the location of the SO ring.
(b) PV map of SO along the sulfur ring (Fig. \ref{fig_irs63_color_maps}).   The white line represents the modeled PV curve of a rotating ring, calculated using Eqs. \ref{eq_ellip_arclength} and \ref{eq_ellip_vproj}. See Sect. \ref{sec_irs63_ring_steamer} for the fitting results.
\label{fig_irs63_pvmaps_ringandS}
}
\end{figure}

\begin{figure}[t]
    \centering
    \includegraphics[width=0.9\linewidth]{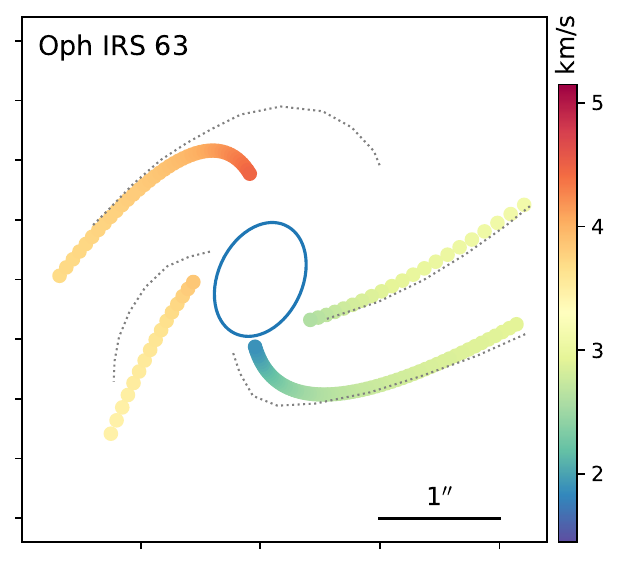}
\caption{Models of the four spiral-like structures in Oph IRS 63 (Figs. \ref{fig_irs63_color_maps} and \ref{fig_irs63_SO_mom12}) generated by free-fall streamers, with colors representing the line-of-sight velocity. Model parameters are provided in Table \ref{tab_fourspiral_models} (Sect. \ref{sec_irs63}). The dotted gray lines indicate the observed spiral-like structures (Fig. \ref{fig_irs63_color_maps}), and the blue ellipse marks the SO ring with a major-axis diameter of 1\arcsec~(130 au). The color map corresponds to that in panel (a) of Fig. \ref{fig_irs63_SO_mom12}. Velocities are scaled by a factor of 2 (from the free-fall values) to approximate the observed velocity range (Fig. \ref{fig_irs63_SO_mom12}), with justification provided in Sect. \ref{sec_irs63_ring_steamer}.
\label{fig_IRS63_model_streamers}
}
\end{figure}

\begin{figure*}
    \centering
    \includegraphics[width=0.8\linewidth]{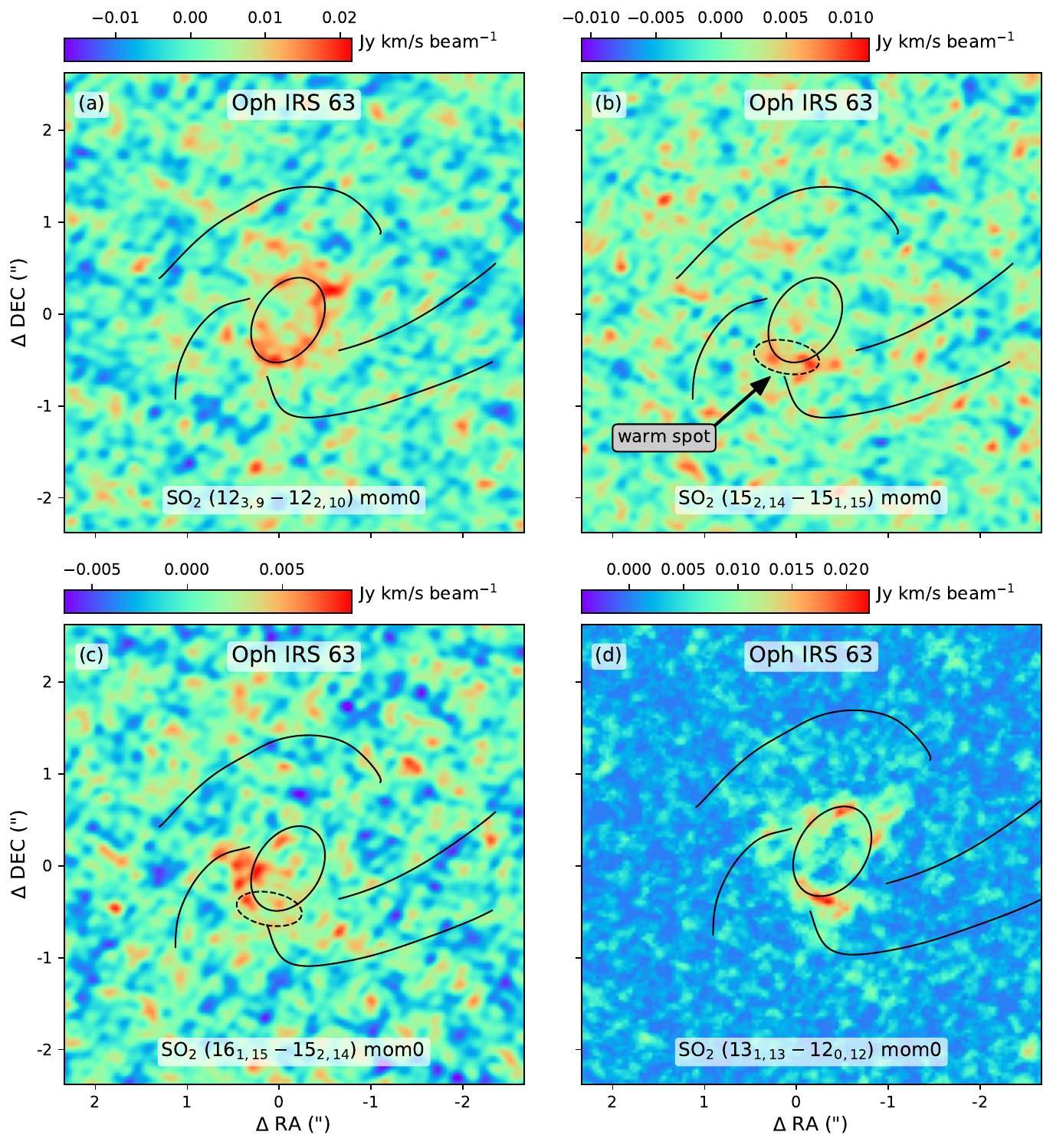}
\caption{Moment 0 maps of Oph IRS 63 for four SO$_2$ transitions (Table~\ref{tab_transitions}).  
In panel (b), the warm SO$_2$ emission spot (Sect.~\ref{sec_warmspot}) is highlighted with a dashed ellipse, which is also shown in panel (c) for comparison.  
The black ellipse and curves denote the SO ring and the spiral structures, respectively (Fig.~\ref{fig_irs63_color_maps}).  
For the moment 0 map of the strongest SO$_2$ line, see Fig.~\ref{fig_irs63_color_maps}. \label{fig_irs63_3lines_so2}}
\end{figure*}


\subsubsection{VLT MUSE} \label{sec_obs_vlt}
We utilized the VLT/MUSE data (PI: M. L. van Gelder) to study the outflow direction of DK Cha (Sect. \ref{sec_dkcha}).
The Multi Unit Spectroscopic Explorer \citep[MUSE;][]{2010SPIE.7735E..08B}, is a panoramic integral-field spectrograph on the Very Large Telescope (VLT). 
The observations were conducted during the night of May 3rd, 2019, in the narrow field mode, with an angular resolution of \mbox{$\sim 0.15\arcsec$} and a resolving power of $R\sim 2500$ at \mbox{$\sim$7000 \AA}. The spectra cover an optical wavelength range of \mbox{4800$-$9300 \AA}. The data were reduced and calibrated using the ESO Reflex pipeline \citep{2013A&A...559A..96F}. Dithering and pipeline calibration were performed to improve the quality of the data.

\subsection{ALMA archival data}
We utilize archival data of DK Cha from the ALMA program 2013.1.00708.S (PI: F. M\'enard). CO (2-1) was observed in this program with a spectral resolution of $\sim 0.16$ km s$^{-1}$ and a synthesized beam resolution of $0.5\arcsec \times 0.25\arcsec$. The observations were conducted on August 27, 2015. We retrieved the spectral cube, processed and cleaned by CASA 5.6.1 as part of the Additional Representative Images for Legacy \citep[ARI-L;][]{2021PASP..133h5001M}, from the ALMA archive for further analysis. 
The 1-$\sigma$ noise level at the spectral resolution of 0.16 km s$^{-1}$ is 35 mJy beam$^{-1}$.

We also utilize archival data of L1527 from ALMA program 2019.A.00034.S (PI: J. Tobin). CO (2-1) was observed in this program with a spectral resolution of $\sim 0.63$ km s$^{-1}$ and a synthesized beam resolution of $0.32\arcsec \times 0.21\arcsec$ \citep{2023ApJ...951...10V}. The observations were conducted on July 3, 2022. These data were used solely to determine the boundaries of CO outflows. The spectral cube from the pipeline products downloaded from the ALMA archive was used without additional processing. The 1-$\sigma$ noise level at a spectral resolution of \mbox{0.63 km s$^{-1}$} is \mbox{2.1 mJy beam$^{-1}$}.

\subsection{Target protostars}\label{sec_sample}
The two ALMA programs (2022.1.01411.S and 2023.1.00592; Sect. \ref{sec_obssetup}) both targeted nearby low-mass protostars, but their samples do not overlap. Some observations of the proposed sources have not been fully completed. This work focuses on three of the sources, selected based on the following criteria: (1) the observations for these sources have been completed; (2) they show resolved signatures of SO and SO$_2$ emission; and (3) their disks have different inclination angles ($\theta_{\rm view}$). 
Table \ref{tab_transitions} lists the ALMA projects and corresponding targets studied in this work.
A brief overview of each target is provided below.

IRAS 16285–2355, commonly known as Oph IRS 63, is a Class I source. The observations are centered at \mbox{RA (J2000) = 16$^{\rm h}$31$^{\rm m}$35.70$^{\rm s}$} and \mbox{Dec (J2000) = $-24^{\rm d}$01$^{\rm m}$29.6$^{\rm s}$}.
Located in L1709 within the Ophiuchus star-forming region \citep{2006AJ....131.2921R}, it is at a distance of 132$\pm$6 pc \citep{2020A&A...633A..51Z}. Oph IRS 63 has a bolometric temperature ($T_{\rm bol}$) of \mbox{348 K} and a bolometric luminosity ($L_{\rm bol}$) of 1.3 $L_\sun$ \citep{2022ApJ...925...12S,2023ApJ...951....8O}. It is one of the brightest Class I protostars, younger than 500\,000 years \citep{2020Natur.586..228S}. Multiple annular substructures have been found in the protoplanetary disk of Oph IRS 63, with a scale of $\sim$150 au and an inclination angle ($\theta_{\rm view}$) of $\sim$45$\degr$ \citep{2020Natur.586..228S}. The SO emission from Oph IRS 63 shows a centrally depleted pattern, with a size comparable to that of the protoplanetary dust disk, surrounded by three spiral structures with a size of several hundred au \citep{2023ApJ...958...98F}. In the literature, the protostellar mass ($M_\star$) of Oph IRS 63 is typically estimated to be \mbox{0.5 $M_\odot$} \citep{2007A&A...461.1037B, 2013A&A...559A..82B, 2023ApJ...958...98F}.

\begin{table}[!t]
\caption{Model parameters of the streamers in Oph IRS 63  (Sect. \ref{sec_irs63}).
\label{tab_fourspiral_models}}
    {\centering
    \begin{tabular}{cccc}
    \hline
        Structure & $\theta_0$ ($\degr$) & $\phi_0$ ($\degr$) & $R_{\rm C}$\\
        S1 &  50 & 130 & 1$\arcsec$ ($\sim 130$ au)\\
        S2 &  80 & 200 & 0.6$\arcsec$ ($\sim 80$ au)\\
        S3 &  125 & -63 & 0.6$\arcsec$ ($\sim 80$ au)\\
        S4 &  80 & -10 & 0.6$\arcsec$ ($\sim 80$ au)\\
         \hline
    \end{tabular}\\
    }
    {\small The  radius of centrifugal barrier 
    ($R_{\rm C}$) of the globally infalling envelope is
    adopted as $30\arcsec$ ($\sim$65 au), the length of the semi-major axis 
    of the SO ring (Fig. \ref{fig_irs63_color_maps}).}
\end{table}

IRAS 12496-7650, also known as DK Cha, is a Herbig Ae star and the most luminous IRAS source in the Chamaeleon II star-forming region, located at a distance of 178 pc \citep{1997A&A...327.1194W}. It is transitioning from Class I to Class II, with an accretion luminosity of 9 $L_\sun$, derived from hydrogen infrared lines \citep{2011A&A...534A..99G,2015A&A...575A..19R}.
Observations of its outflows \citep{2015A&A...576A.109Y,2023ApJ...945...63H} suggest that DK Cha is a nearly face-on system with a small inclination angle. However, no ALMA long-baseline observations\footnote{\url{https://almascience.eso.org/aq}} with a resolution better than 0.1$\arcsec$ have been obtained, and its disk has not been resolved. Consequently, there remains some uncertainty regarding the inclination angle of its disk. The observations of DK Cha are centered at \mbox{RA (J2000) = 12$^{\rm h}$53$^{\rm m}$17.07$^{\rm s}$}, \mbox{Dec (J2000) = $-$77$^{\rm d}$07$^{\rm m}$10.8$^{\rm s}$}.

IRAS 04368+2557, also known as L1527, is a Class 0 low-mass protostar embedded in a protostellar core within the Taurus molecular cloud, located at a distance of 144 pc \citep{2007ApJ...671.1813T,2012Natur.492...83T}.
It features a nearly edge-on disk \citep[$\theta_{\rm view}\sim 90\degr$;][]{2014Natur.507...78S,2018ApJ...861...91H,2023ApJ...951...10V,2024ApJ...966..207Z}.
The central mass of L1527 was initially estimated to be $\sim 0.2$ $M_\sun$ \citep{2012Natur.492...83T,2016MNRAS.463.3563T}, but recent observations suggest a value of \mbox{$\sim 0.5$ $M_\sun$} \citep{2023ApJ...951...10V}.
The SO emission from L1527 exhibits a structure resembling parallel slabs \citep{2017MNRAS.467L..76S}, which has been interpreted as a ring formed by the centrifugal barrier of infalling gas \citep{2014Natur.507...78S,2017MNRAS.467L..76S}.
The target center of L1527 is located at \mbox{RA (J2000) = 04$^{\rm h}$39$^{\rm m}$53.87$^{\rm s}$} and \mbox{DEC (J2000) = $+$26$^{\rm d}$03$^{\rm m}$09.5$^{\rm s}$}.

\section{Streamers and accretion shock traced by SO/SO$_2$}\label{sec_obsresults}
\subsection{Oph IRS 63}\label{sec_irs63}
In Oph IRS 63, a low-mass protostar, distinct signatures of accretion shocks have been observed, including ring-like structures and spiral-like streamers in the SO and SO$_2$ emission. These features provide compelling evidence for the presence of shocks at/around the disk-envelope interface during the early stages of star formation.
To better understand these shock-related structures, we first examine the detailed morphology and kinematics of the SO emission, beginning with the ring-like component.
\subsubsection{SO ring}
\label{sec_irs63_ring_steamer}
Among the two observed SO lines (Table \ref{tab_transitions}), only SO ($5_6-4_5$) is detected in Oph IRS 63. SO ($5_6-4_5$) displays a ring-like structure with multiple spiral-like features (see panels (a) and (b) of Fig. \ref{fig_irs63_color_maps}). The morphology is similar to that revealed by \citet{2023ApJ...958...98F} using the SO ($6_5-5_4$) emission at a resolution of 0.3--0.4\arcsec. Due to the higher resolution of this study ($\sim 0.2\arcsec$), we are able to observe the SO ring in greater detail and identify that the inner edge of the SO ring resembles an elliptical shape. The elliptical inner boundary of SO has a size comparable to that of the second (outer) dust ring of the protoplanetary disk, as revealed by \citet{2020Natur.586..228S} using the ALMA long-baseline data (see also Fig. \ref{fig_irs63_color_maps}). 
We fit the ring-like SO emission using an elliptical ring model, following the algorithm described in Appendix~\ref{sec_fit_ring}. The azimuthal angle of the major axis ($\phi_{\rm view}$) and the inclination angle ($\theta_{\rm view}$) are fitted to be $53\pm1^\circ$ and $40.5\pm1^\circ$, respectively. The fitted value of $\theta_{\rm view}$ is close to that reported by \citet{2020Natur.586..228S} ($45^\circ$). The small discrepancy may result from deviations from a perfect elliptical shape, or from a possible misalignment between the disk and the SO ring.
In this work, we define $\phi_{\rm view}=0$ for a structure whose major axis lies in the east-to-west direction, with $\phi_{\rm view}$ increasing as the major axis rotates counterclockwise. 
Note that in some studies, the position angle (PA) is defined starting from the north direction.
During the follow-up fitting, we fixed $\theta_{\rm view}$ to 45\degr.

The moment 1 map of SO (Fig.~\ref{fig_irs63_SO_mom12}(a)) reveals an hourglass-shaped structure composed of blue- and redshifted components aligned along the major axis of the ring.  
In the moment 2 map (Fig.~\ref{fig_irs63_SO_mom12}(b)), an S-shaped broad-line feature is apparent (see also Fig. \ref{fig_irs63_Smark} in Appendix \ref{sec_ectraimage}).  
This S-shaped structure corresponds to the southeast–northwest margin of the hourglass seen in the moment 1 map and exhibits a steeper velocity gradient than the southwest–northeast margin.  
The position–velocity (PV) diagram extracted along the S-shaped feature (panel (a) of Fig.~\ref{fig_irs63_pvmaps_ringandS}) shows distinct velocity components, which we will discuss further in Sect.~\ref{sec_beyonddis}.

Here, we focus on analyzing the dynamics of the SO ring itself.  
We fit the velocity distribution along the ring (panel (a) of Fig.~\ref{fig_irs63_SO_mom12}) assuming pure rotation, following Eqs.~\ref{eq_ellip_arclength} and \ref{eq_ellip_vproj}.  
The fitting results are shown in panel (b) of Fig.~\ref{fig_irs63_pvmaps_ringandS}.  
The rotational velocity ($V_{\rm rot}$) and systemic velocity ($V_{\rm sys}$) are fitted to be approximately 2.2 km~s$^{-1}$ and 3 km~s$^{-1}$, respectively.  
Assuming Keplerian rotation at a radius of 65~au, this corresponds to a central mass of $\sim 0.35\ M_\sun$, slightly smaller than the $\sim 0.5\ M_\sun$ reported in the literature (Sect.~\ref{sec_sample}).

\subsubsection{SO spiral-like streamers of different types}\label{sec_streamer_irs63}
While the ring-like SO structure reveals rotation at the disk-envelope interface (Sect. \ref{sec_irs63_ring_steamer}), additional substructures are present in the form of spiral-like extensions. We now turn to these features, which appear to trace dynamically distinct infalling motions.
Four spiral-like structures, denoted as S1 to S4, are identified in the SO emission, with their feet located around the SO ring and their tails extending outward. Among these, S1 to S3 correspond to the three arms reported by \citet{2023ApJ...958...98F}, while the S4 structure was not resolved in their work due to slightly lower resolution and sensitivity. In the moment 0 map of SO ($5_6$--$4_5$) shown in Fig.~\ref{fig_irs63_color_maps}, the western tail and the foot of S4 exhibit an overall S/N exceeding 4, although the emission at the connection between them is weaker (S/N~$\lesssim$~3). Given the symmetric distribution of the four feet S1--S4 (Fig.~\ref{fig:ringfit}), we consider the presence of S4 to be plausible.

To interpret the origin of the spiral structures, we model them as gas streamers undergoing ballistic free-fall toward the disk. Their three-dimensional trajectories are computed and then projected onto the plane of the sky, following the procedure outlined in Appendix~\ref{sec_parabolic_trajectory}. 
The free-falling gas, characterized by zero mechanical energy and non-zero specific angular momentum, follows a parabolic trajectory in 3D space. This trajectory is parameterized by the initial polar angle ($\theta_0$), azimuthal angle ($\phi_0$), and centrifugal radius ($R_{\rm C}$) within the disk coordinate system.
For a free-fall stream with a given specific angular momentum $J$, the centrifugal radius $R_{\rm C}$ represents the radius of an equivalent circular orbit conserving the same angular momentum. Geometrically, $R_{\rm C}$ also corresponds to twice the distance between the focus and the vertex of the parabola. Streams originating from different directions are expected to intersect at or beyond $R_{\rm C}$, preventing material from freely falling all the way to the vertex of the parabolic path.
Additionally, $R_{\rm C}$ equals the distance between the parabola’s focus and its directrix (see Sect.~\ref{sec_cmu} for details).
When projected onto a two-dimensional plane defined by viewing angles $\theta_{\rm view}$ and $\phi_{\rm view}$, the trajectory remains parabolic, although the central star may no longer lie exactly at the focus (see Sect.~\ref{sec_model} and Appendix~\ref{sec_parabolic_trajectory}).

To model the trajectories of the observed spiral-like features, we adopt a simplified, hand-drawn, illustrative approach aimed at qualitatively matching the observed morphology. As a first step, we assess whether the observed structures can be explained by free-fall streamers confined strictly to the disk midplane. According to \mbox{Eq.~\ref{eq_phi_incre}} in \mbox{Appendix~\ref{sec_eq_details}}, such midplane-confined streamers (corresponding to $\theta_0 = \pi/2$) can span an azimuthal angle on the plane of the sky ($\theta_{\rm proj}$) of less than $110^\circ$. However, some features—such as S1 in Fig.~\ref{fig_irs63_color_maps}—have $\Delta \theta_{\rm proj} > 110^\circ$, indicating that their trajectories cannot be explained by midplane infall alone. Motivated by this, we fix the viewing angles $\theta_{\rm view}$ and $\phi_{\rm view}$ to match the inclination and azimuth of the SO ring (and thus the protoplanetary disk; \citealt{2020Natur.586..228S}), and manually adjust only the model parameters—$\theta_0$, $\phi_0$, and $R_{\rm C}$—to better reproduce the observed spiral shapes. The uncertainty of $\theta_0$ is approximately $10^\circ$.

We define the off-midplane angle as $\theta_0^{\rm off\,disk} = |90^\circ - \theta_0|$, which quantifies the deviation of a streamer's initial direction from the disk midplane. In this convention, $\theta_0^{\rm off\,disk} = 0^\circ$ corresponds to a trajectory lying exactly in the midplane, while $\theta_0^{\rm off\,disk} = 90^\circ$ corresponds to infall along the polar axis. For S1 and S4, we find $\theta_0^{\rm off\,disk} \gtrsim 30^\circ$, whereas for S2, it is approximately $10^\circ$. The off-midplane angle for S4 is not well constrained; therefore, given the morphological similarity between S1 and S3, we adopt the $\theta_0^{\rm off\,disk}$ value of S3 for S4. The modeled trajectories are illustrated in Fig.~\ref{fig_IRS63_model_streamers}, and their parameters are summarized in Table~\ref{tab_fourspiral_models}.

The above fitting assumes a central stellar mass of 0.35~$M_\sun$ (Sect. \ref{sec_irs63_ring_steamer}). Even after accounting for projection effects, the modeled free-fall velocities are found to be slightly higher than the observed values. To reconcile this discrepancy, we manually scale down the free-fall velocities by a factor of approximately 1.5. This adjustment is justifiable, as the observed velocities correspond to line-center measurements rather than the full velocity span. As demonstrated by \citet{2023ApJ...958...98F}, central stellar masses derived from fitting the ridge versus the outer boundary of emission in position-velocity diagrams can differ by a factor of $\sim 3$. Furthermore, the free-fall assumption may not be strictly applicable to all streamers, particularly in cases where interactions with the inner envelope alter their dynamics (see Sects.~\ref{sec_irs63_envelope} and \ref{sec_streamer_onoff}).


Based on these results, we classify the spiral-like structures as streamers, supported by several key considerations. First, their large off-midplane angles ($\theta_0^{\rm off\,disk}$) indicate significant deviation from the disk midplane. This distinguishes them from typical spiral arms, which are generally confined to the midplane and shaped primarily by disk or inner envelope dynamics (Sect.~\ref{sec_model}). Second, two distinct SO emission spots are observed at the bases of S1 and S3—located near the endpoints of the SO ring’s major axis. These spots are clearly resolved in our data (Fig.~\ref{fig_irs63_color_maps}) and may mark the locations where the associated off-midplane streamers appear to intersect the disk surface. Finally, all identified streamers originate from beyond the centrifugal radius (see Sect.~\ref{sec_cmu} for the definition of centrifugal radius), where they likely interact with the rotating and infalling inner envelope (see Sect.~\ref{sec_irs63_envelope} for the observational evidence of the inner envelope). These interactions may play an important dynamical role, producing localized heating and chemical enhancement where streamers impact the disk or envelope surface. In the next subsection, we explore such evidence by examining SO$_2$ emission associated with the base of one prominent streamer (Sect.~\ref{sec_warmspot}).

To account for the diversity in observed morphologies and infall geometries, we further classify the streamers into two distinct types: in-midplane and off-midplane, corresponding to relatively small (e.g., $\lesssim 30^\circ$) and large (e.g., $\gtrsim 30^\circ$) values of $\theta_0^{\rm off\,disk}$, respectively. Additional evidence for both in-midplane and off-midplane streamers will be presented for other observational targets in Sect.~\ref{sec_dkcha}. Their differing interactions with the disk and envelope structures will be discussed in our modeling framework in Sect.~\ref{sec_model}.

\begin{figure}[t]
    \centering
    \includegraphics[width=0.99\linewidth]{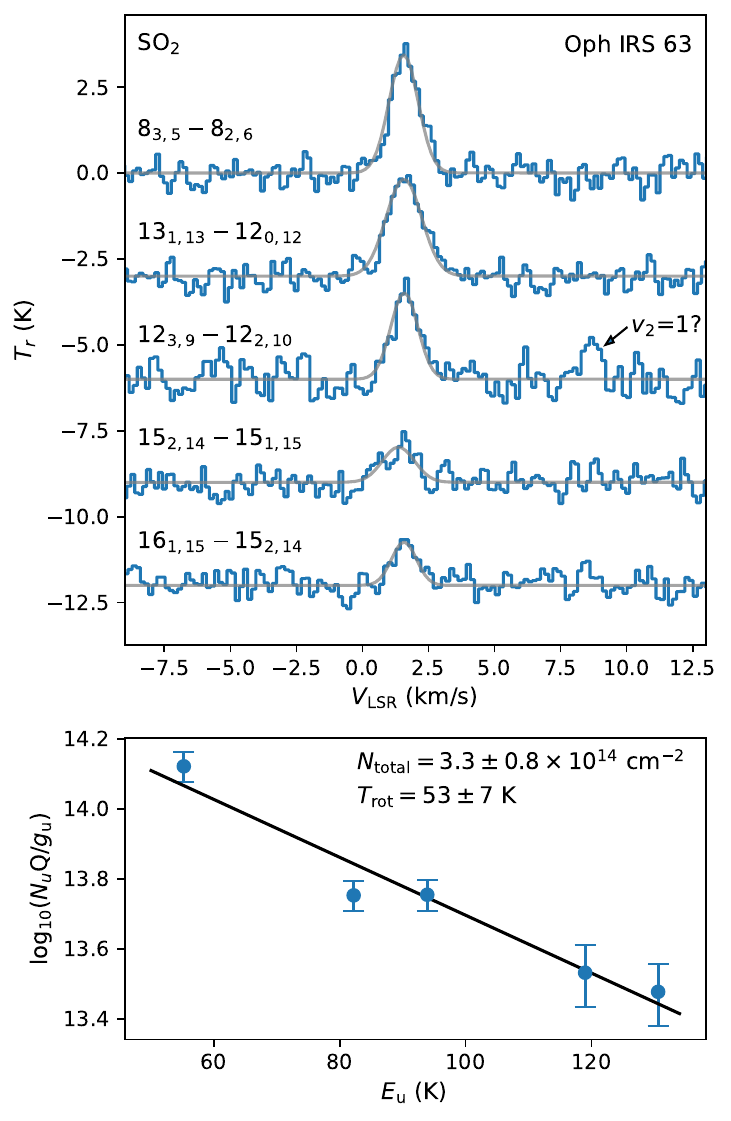}
\caption{Upper: SO$_2$ spectra of the warm spot in Oph IRS 63, enclosed by a black dashed ellipse in Fig. \ref{fig_irs63_3lines_so2}. Lower: Rotational diagram of SO$_2$ in the warm spot. \label{fig_so2_spectra}
}
\end{figure}

\begin{figure}[!t]
    \centering
    \includegraphics[width=0.99\linewidth]{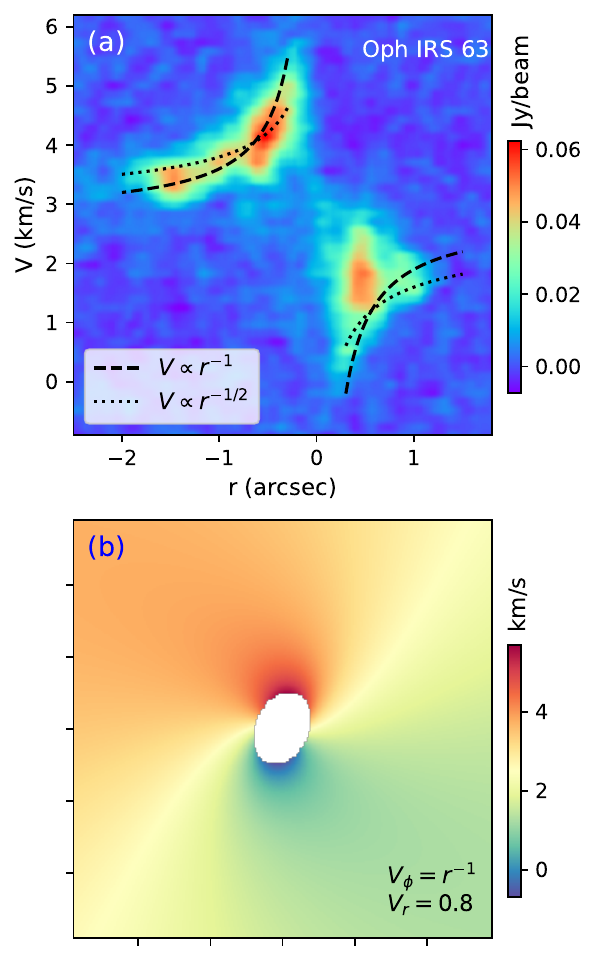}
\caption{(a) The position-velocity (PV) map of SO in Oph IRS 63 along the major axis of the SO ring marked in Fig. \ref{fig_irs63_color_maps}. The dashed and dotted black lines represent the fits to the velocity distribution of $V_\phi$ with power-law indices of $-1$ (the free-fall case with non-zero angular momentum) and $-0.5$ (the Keplerian case), respectively (see Sect. \ref{sec_shockedsurface} for details).  (b) The modeled velocity of an inner disk-like envelope (in a simplified 2D version) with its projection parameters ($\theta_{\rm view}$ and $\phi_{\rm view}$) identical to those of Oph IRS 63. The radial ($v_r$) and azimuthal ($V_\phi$) velocities are given by the formulas in the legend, following Eqs. \ref{eq_mod_vr} and \ref{eq_mod_vphi}, which describe the possible motion of disk-like inner envelopes (see Sect. \ref{sec_shockedsurface} for details). 
Note that the model is scale-free, and we have adjusted the velocity range and central velocity for better agreement with the observations (panel (a) of Fig. \ref{fig_irs63_SO_mom12}).
\label{fig_pv_irs63}
}

\end{figure}

\subsubsection{SO$_2$ ring and warm spot}\label{sec_warmspot}
All five SO$_2$ lines (Table~\ref{tab_transitions}) are detected toward Oph IRS 63 (Figs.~\ref{fig_irs63_color_maps} and \ref{fig_irs63_3lines_so2}). 
Among these, the SO$_2$ ($8_{3,5} - 8_{2,6}$) transition has the lowest upper-level energy and the highest S/N. Its moment 0 map shows an S/N of approximately 6 (Fig.~\ref{fig_irs63_color_maps}).
The emission from SO$_2$ ($13_{1,13} - 12_{0,12}$) (Fig.~\ref{fig_irs63_3lines_so2}) resembles that of SO$_2$ ($8_{3,5} - 8_{2,6}$) (right panel of Fig.~\ref{fig_irs63_color_maps}), although with a slightly lower S/N of $\sim$5 in its moment 0 map. As with SO, the moment 0 maps of these two SO$_2$ lines reveal a ring-like structure (Figs.~\ref{fig_irs63_color_maps} and \ref{fig_irs63_3lines_so2}). The SO$_2$ ring appears to be comparable in size to, or possibly slightly smaller than, the SO ring. However, due to the current resolution and sensitivity limits, we cannot definitively distinguish between the SO and SO$_2$ rings. We therefore refer to them collectively as the sulfur-oxide rings of Oph IRS 63.

The feet of the streamers are located near the sulfur-oxide ring (Sect.~\ref{sec_streamer_irs63}). At these feet, the emission of \mbox{SO$_2$ ($8_{3,5}-8_{2,6}$)} and SO$_2$ ($12_{3,9}-12_{2,10}$) is typically enhanced (Figs.~\ref{fig_irs63_color_maps} and \ref{fig_irs63_3lines_so2}). In particular, an emission spot of \mbox{SO$_2$ ($15_{2,14}-15_{1,15}$)} is only marginally detected at the foot of S3 (see Fig.~\ref{fig_irs63_3lines_so2}), with a S/N of $\sim$3 in the moment 0 map. However, the average spectrum of SO$_2$ ($15_{2,14} - 15_{1,15}$) at this spot shows a clear detection, with a S/N greater than 5 (Fig.~\ref{fig_so2_spectra}). This SO$_2$ spot is spatially close to the SO spot in the same region.
The SO$_2$ emission appears more extended than the SO spot and seems to stretch toward the bases of S2 and S3 (Fig.~\ref{fig_irs63_3lines_so2}), though this interpretation remains tentative due to the relatively low S/N. Notably, the \mbox{SO$_2$ ($15_{2,14}-15_{1,15}$)} transition has a high upper-level energy of $E_u \sim 120$\,K (Table~\ref{tab_transitions}), suggesting that the spot at the foot of S3 traces gas that is warmer and/or denser than the average conditions in the sulfur-oxide ring. This may result from the interaction between S3 and the inner envelope or disk near the centrifugal barrier. The large off-midplane angle $\theta_0^{\rm off\,disk}$ of S3 allows it to fall directly onto the midplane in the vicinity of the sulfur-oxide ring (Sect.~\ref{sec_streamer_onoff}), potentially generating strong shocks that excite and produce warm SO$_2$ emission.
These findings support the interpretation of the spiral-like structures as gas streamers associated with the inner envelope, where their infall motions play a dynamically and chemically significant role near the centrifugal barrier. Our results suggest that the sulfur-oxide ring is likely induced by shocks around the centrifugal barrier, with SO$_2$ serving as a more sensitive tracer of strong shocks than SO \citep{2021A&A...653A.159V}.

To further test this interpretation and assess the thermochemical state of the gas, we analyze the SO$_2$ line profiles and construct a rotational diagram of the warm spot (Fig.~\ref{fig_so2_spectra}). The excitation temperature ($T_{\rm ex}$) and column density ($N$) of SO$_2$ are determined to be 53\,K and $3.3 \times 10^{14}$\,cm$^{-2}$, respectively, using the method described in Appendix~\ref{sec_cal_rtdiagram}. These results confirm that the SO$_2$ spot is indeed warm, with a gas temperature $\gtrsim 50$\,K—typical of post-shock molecular gas \citep{2015A&A...578A..63F,2021A&A...653A.159V}.
The SO spectrum at the warm spot exhibits a line width of 1.5\,km\,s$^{-1}$, a peak brightness temperature of 15\,K, and an integrated intensity of 23.4\,K\,km\,s$^{-1}$. The main uncertainty in the derived column densities stems from the excitation temperature, which is estimated to be accurate within 20\% (Fig.~\ref{fig_so2_spectra}). Adopting the same $T_{\rm ex}$ as for SO$_2$, the SO column density is calculated to be $5 \times 10^{14}$\,cm$^{-2}$, corrected for optical depth effects ($\tau \lesssim 0.3$; Appendix~\ref{sec_cal_rtdiagram}). The resulting column density ratio $N$(SO$_2$)/$N$(SO) is approximately 0.6—significantly higher than the value of 0.03 predicted by the fiducial shock model of \citet{2021A&A...653A.159V}.
Both SO and SO$_2$ can be efficiently formed through reactions of atomic S with OH in high-velocity ($>$4\,km\,s$^{-1}$) shocks in moderately dense environments \citep{2021A&A...653A.159V}. To account for the observed high SO$_2$/SO ratio, a shock velocity exceeding 4\,km\,s$^{-1}$ is required, which is consistent with the velocity range observed in the SO ring (see color bar in panel~(a) of Fig.~\ref{fig_irs63_SO_mom12}). A streamer impacting the centrifugal barrier can generate such a strong shock, enhancing SO$_2$ production. In contrast, streamers that co-rotate with the inner envelope are less likely to experience this level of shock interaction (Sect.~\ref{sec_shockedsurface}). We therefore conclude that the observed SO$_2$ emission at the foot of S3 is likely induced by strong accretion shocks.

Knowing the excitation temperature allows us to estimate the density of the shocked gas. Assuming a warm spot size of 0.2$\arcsec$ (or $L_{\rm spot} \sim 26$\,au) and a post-shock SO abundance ($X_{\rm SO}$) of $10^{-8}$ \citep{2021A&A...653A.159V}, the number density ($n_{\rm H_2}$) at the warm spot can be estimated as:
\begin{equation}
n_{\rm H_2} = \frac{N_{\rm SO}}{X_{\rm SO} L_{\rm spot}} \sim 10^8\ {\rm cm}^{-3}. \label{eq_nh2_spot}
\end{equation}
Free-fall envelope models—such as the classical CMU model \citep{1976ApJ...210..377U,1981Icar...48..353C} and the constant-$J$ model (see Sect.~\ref{sec_hardsurface} for a brief introduction)—predict a divergent density at the centrifugal radius ($R_{\rm C}$) due to the neglect of pressure (e.g., Eq.~\ref{eq_rhoenv}). Given that both the radial thickness of the SO ring and the size of the warm spot are approximately $0.4\,R_{\rm C}$, we adopt the density at $1.4\,R_{\rm C}$ from the constant-$J$ model as a representative value for the inner envelope. Assuming an accretion rate of $10^{-6}$\,$M_\sun$\,yr$^{-1}$ \citep{2023ApJ...958...98F}, this yields a typical density of $\sim 10^7$\,cm$^{-3}$ for the inner envelope—slightly lower than the estimated density at the warm spot (Eq.~\ref{eq_nh2_spot}). This discrepancy may arise from (1) a higher density in the streamers compared to the surrounding envelope, and/or (2) accumulation of material in the post-shock gas.

\begin{figure*}
    \includegraphics[width=0.99\linewidth]{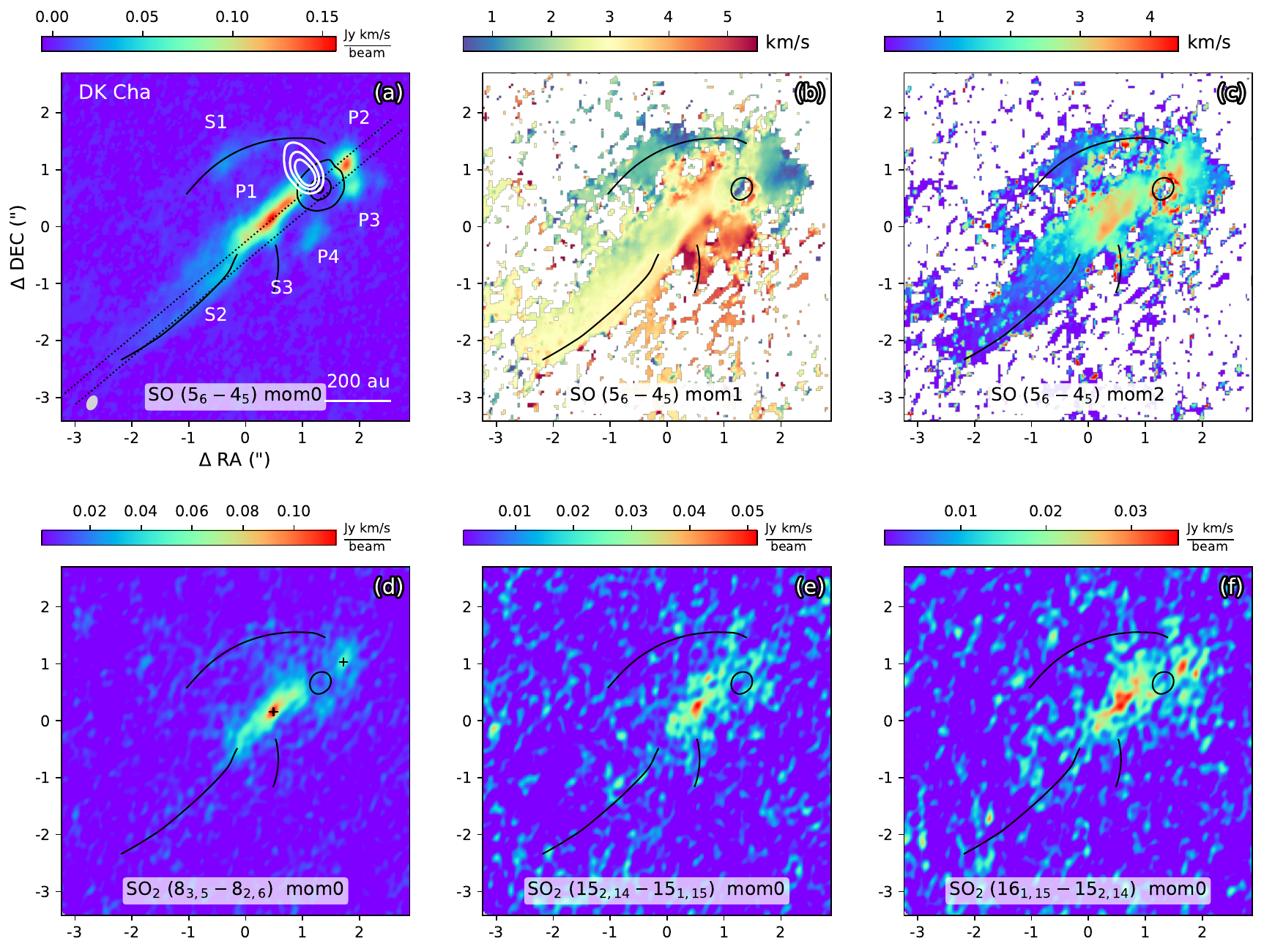}
    \caption{The moment maps of SO and SO$_2$ for DK Cha. The transition quantum labels are provided at the bottom of each panel, and the beam is indicated in the lower left corner of the first panel. In panel (a), the black contours represent the ALMA continuum emission at levels of \mbox{[0.01, 0.5, 0.9]} times the peak value (0.16 Jy beam$^{-1}$). 
    In the other panels, only the 0.5 times the peak value contour of the ALMA continuum is shown. The white contours in the first panel outline the north-east lobe of the outflow, as traced by [S {\sc ii}] 6731 \r{A} using VLT/MUSE. This outflow has been smoothed to match the resolution of the ALMA data. The white contour levels are [0.3, 0.5, 0.7] times the peak value. Substructures identified in the moment 0 map of SO are labeled in the upper-left panel, with three streamers (S1 to S3) marked by black curves. 
    For an illustration of additional substructures, their names, and connections, refer to the schematic image in Fig. \ref{fig_dkcha_cartoon}. 
    As in panel (b) of Fig. \ref{fig_irs63_SO_mom12}, the moment 2 map of SO in this figure has been multiplied by a factor of $\sqrt{8\ln(2)}$.
    In panel (d), the two black crosses indicate the positions used to extract the rotation diagrams shown in Fig.~\ref{fig_dkcha_warmspots}.
    The center of the shown region is located at \mbox{RA (J2000) = 12$^{\rm h}$53$^{\rm m}$17.45$^{\rm s}$}, \mbox{DEC (J2000) = $-$77$^{\rm d}$07$^{\rm m}$11.5$^{\rm s}$}, which does not coincide with the protostar's location.
     \label{fig_moms_dkcha}
     }
\end{figure*}

\subsubsection{Inner envelope motion}\label{sec_irs63_envelope}

In addition to the SO rings and spiral-like structures, weak extended SO emission is detected outside the ring. Along the minor axis, the northeast region shows a slight redshift, while the southwest region is slightly blueshifted (panel (b) of Fig.~\ref{fig_irs63_SO_mom12}). Because pure rotational motion would produce zero projected velocity along the minor axis, this pattern indicates the presence of both rotational and infalling motions. Given that this extended emission lies primarily outside the centrifugal barrier, we interpret it as tracing the surface of a disk-like envelope confined by an accretion shock. This interpretation is developed based on our combined observational results and is presented in detail in Sect.~\ref{sec_model}, particularly in Sect.~\ref{sec_shockedsurface}. In other words, we interpret this emission as likely originating from the rotating and infalling inner envelope.


To support the above hypothesis, we first examine the rotational motion of the extended structure. The rotational and infalling velocity components may be coupled; however, the projection of the infalling velocity component along the major axis of the disk and inner envelope is expected to be zero (assuming negligible envelope flare). Therefore, the position-velocity (PV) map of SO along the major axis of the SO ring (panel (a) of Fig.~\ref{fig_pv_irs63}) primarily reflects the rotational velocity component.
The rotational velocity of the inner envelope is expected to follow a power-law with radius, characterized by an index $\beta$ that can range from 0 (free fall with zero angular momentum) to $-1/2$ (Keplerian rotation), or even $-1$ (free fall with conserved angular momentum). This framework is further detailed in Eq.~\ref{eq_mod_vphi} and Sect.~\ref{sec_shockedsurface}. If the motion of the post-shock surface traced by the extended SO emission is fully governed by the underlying gas flows, one would anticipate a velocity index close to $\beta = -1$.
However, given the current sensitivity, we cannot conclusively distinguish between $\beta = -1$ and $\beta = -1/2$. Thus, while both pure rotational and rotating–infalling motions remain consistent with the velocity distribution along the major axis, our combined kinematic analysis confirms the presence of infalling motion in the extended structure.

Furthermore, we modeled the 2D distribution of the projected velocity for a rotating and infalling inner envelope (see panel (b) of Fig.~\ref{fig_pv_irs63}), using Eqs.~\ref{eq_mod_vr} and \ref{eq_mod_vphi}. The rotation follows a power-law with an index ($\beta$) of $-1$, while the infall velocity is fixed. We adjusted the ratio of these velocities to match the observations. The modeled velocity distribution reproduces key features of the observed data (Fig.~\ref{fig_irs63_SO_mom12}), including: (1) the slight offset between the major axis of the SO ring (represented by the blank ellipse in panel (b) of Fig.~\ref{fig_pv_irs63}) and the symmetry axis of the moment-1 map; and (2) the S-shaped structure in the moment-1 map, extending from the lower left to the upper right of panel (b) in Fig.~\ref{fig_pv_irs63} (see panel (a) of Fig.~\ref{fig_irs63_SO_mom12} for the observed S-shape). This confirms that the extended SO emission outside the ring can be explained by emission from a disk-like inner envelope, potentially originating from its surface, with its rotating–infalling motion tracing the continuation of the free-fall streams with non-zero angular momentum.

These modeling results provide a physical basis for interpreting the structure and dynamics of the inner envelope, which plays a critical role in mediating accretion from large scales. Building on this, we now turn to the interaction between the streamers and the envelope. The inner envelope offers a valuable window into accretion processes near the disk and the behavior of gas streamers. Conceptually, off-midplane streamers are expected to have a strong dynamical impact on the envelope at their collision sites, while in-midplane streamers may gradually merge with the inner envelope and effectively serve as part of it. This distinction provides a qualitative framework for understanding the varying shock activity associated with different types of streamers. A more detailed discussion of these interactions is developed in the context of the disk–envelope framework in Sects.~\ref{sec_shockedsurface} and \ref{sec_streamer_onoff}.

\begin{figure}
    \centering
    \includegraphics[width=0.99\linewidth]{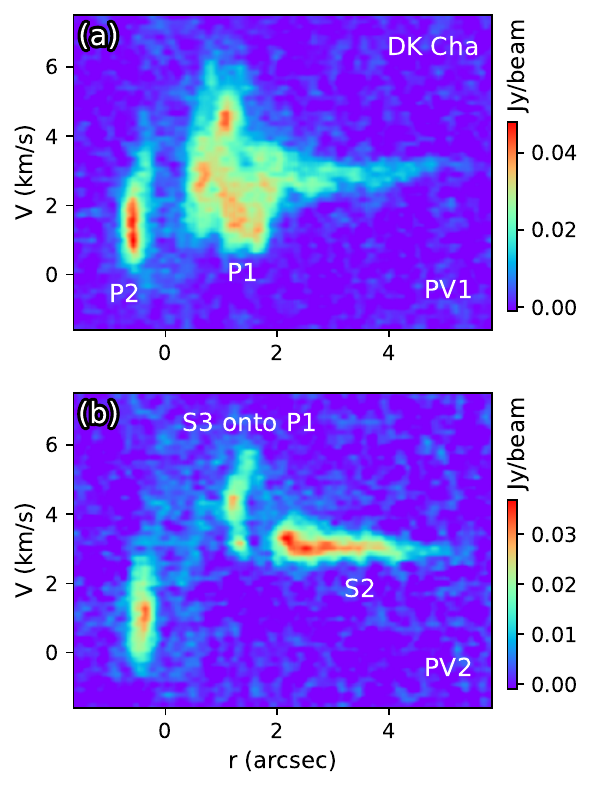}
    \caption{The position-velocity (PV) maps along the two dotted lines in panel (a) of Fig.~\ref{fig_moms_dkcha}. 
    Panels (a) and (b) correspond to the northern and southern lines in Fig.~\ref{fig_moms_dkcha}, respectively. 
    Several gas components identified in Fig.~\ref{fig_moms_dkcha} are visible in the PV maps, and are labeled accordingly.\label{fig_dkcha_pv}}
\end{figure}

\subsubsection{Tentative features in the SO$_2$ line profiles}
At the warm spot, the SO$_2$ spectra with higher $E_{\rm u}$ appear to be more blueward skewed (Fig. \ref{fig_so2_spectra}). \mbox{SO$_2$ ($15_{2,14}-15_{1,15}$)} exhibits the most prominent characteristic with a blue shoulder, albeit with a S/N of only $\sim 2.5$. We tentatively interpret this as evidence that S3 has a large $\theta_0^{\rm off\ disk}$ and impacts the midplane with a blueshifted velocity (see Fig. \ref{fig_IRS63_model_streamers}). Near the emission line of SO$_2$ ($12_{3,9}-12_{2,10})$, there is a weak ($2\sigma$) line feature at \mbox{$\sim 8.5$ km s$^{-1}$} with respect to the rest frequency of \mbox{SO$_2$ ($12_{3,9}-12_{2,10}$}). This feature has a frequency close to the rest frequency of the transition $26_{3,23}-25_{4,22}$ of \mbox{SO$_2$ $v_2=1$}. 
This transition has a very high $E_{\rm u}$ value of $>1117$ K. The ground-level energy of SO$_2$ $v_2=1$ is $\sim745$ K \citep{2005JMoSp.232..213M}, and this transition has an $E_{\rm u}$ of 372 K with respect to the ground level of SO$_2$ $v_2=1$. 
If the emission line of this transition has a velocity shift similar to that of \mbox{SO$_2$ ($12_{3,9}-12_{2,10}$)}, its peak should appear at $\sim 10$ km s$^{-1}$. 
The location of this line feature on the velocity axis is consistent with the blueward skewing trend observed for the high-$E_{\rm u}$ lines of SO$_2$ at the warm spot. 
Considering that the $T_{\rm ex}$ given by the rotational diagram of SO$_2$ (Fig. \ref{fig_so2_spectra}) is only a few tens of Kelvin, this transition would be unlikely to be excited if it originates from the same gas component traced by the relatively low-$E_{\rm u}$ lines. One possibility is the presence of a hotter gas component that allows for the excitation of vibrational states of SO$_2$ $v_2=1$, via processes such as infrared pumping \citep{2024A&A...682A..78V}. 
While we cannot entirely rule out this scenario, we consider it more likely that this line feature is a random observational spike, given its low S/N in Oph IRS 63 and the lack of a corresponding detection in DK Cha, an object with much stronger emission of sulfur-bearing molecules and warmer SO$_2$ spots (Sect. \ref{sec_dkcha}).

\begin{figure}[t]
    \centering
    \includegraphics[width=0.99\linewidth]{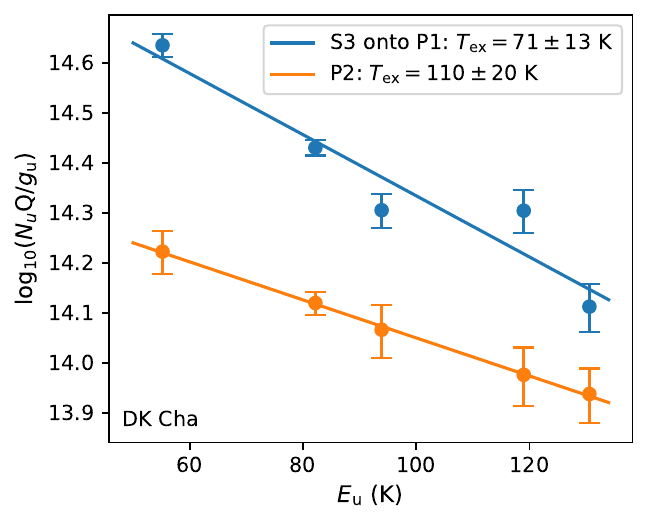}
    \caption{The rotation diagrams of the SO$_2$ emission of the two warm spots (P2 and $\rm S3\perp P1$) of DK Cha (Fig. \ref{fig_moms_dkcha}).
    The emission is extracted at the positions marked by black crosses in Fig.~\ref{fig_moms_dkcha}.}
    \label{fig_dkcha_warmspots}
\end{figure}

\subsection{DK Cha}\label{sec_dkcha}
While clear signatures of accretion shocks, such as ring-like features and spiral streamers in SO and SO$_2$ emission, were observed in Oph IRS 63, the situation for DK Cha, an intermediate-mass protostar, is less straightforward. Nevertheless, its likely more face-on orientation and the limited resolution of the available data still provide valuable insights into how shock features might manifest in a different mass regime.

\subsubsection{Inclination angle of the disk continuum} \label{sec_dkcha_SandP}
There are no ALMA long-baseline data (with a beam resolution better than $0.1\arcsec$) available for this object in the archive. Fig. \ref{fig_moms_dkcha} shows the moment maps of SO and SO$_2$ for DK Cha, overlaid with its dust continuum at a resolution of $\sim 0.2\arcsec$. Although our data do not fully resolve the disk, the nearly round morphology of the continuum emission suggests that DK Cha has a nearly face-on disk. 
The 2D Gaussian fitting of the continuum gives a major-axis FWHM of \mbox{$0.44\arcsec \pm 0.01\arcsec$}, a minor-axis FWHM of \mbox{$0.38\arcsec \pm 0.11\arcsec$}, and a position angle ($\phi_{\rm view}$) of \mbox{$43^\circ\pm 10^\circ$}. The values for the synthesis beam are $0.22\arcsec$, $0.15\arcsec$, and $65^\circ$, respectively. The beam-deconvolved\footnote{The deconvolution of 2D Gaussians can be performed using the Python package \protect\textit{radio\_beam} (\url{https://radio-beam.readthedocs.io/en/latest/}).} major-to-minor axis ratio of the continuum emission gives a $\theta_{\rm view}$ of $\sim 20^\circ$
. This result is consistent with the findings from the analysis of the multi-cone outflows of DK Cha by \citet{2023ApJ...945...63H}, who reported a $\theta_{\rm view}$ of $19^\circ$. The \mbox{[S {\sc ii}] 6731 \r{A}} data from VLA MUSE (Sect. \ref{sec_obs_vlt}) show an outflow lobe in the north-east direction of DK Cha (panel (a) of Fig. \ref{fig_moms_dkcha}). This is likely the blueshifted lobe, while the redshifted one is not visible due to extinction. This confirms that the disk of DK Cha is nearly face-on with a small $\theta_{\rm view}$. Therefore, we adopt a $\theta_{\rm view}$ of $20^\circ$ for the disk of DK Cha in this work. Similar to Oph IRS 63, the emission of SO and SO$_2$ is absent in the disk region of DK Cha, as traced by the dust continuum. The nearly face-on geometry of DK Cha and the absence of SO and SO$_2$ emission within the disk region, similar to Oph IRS 63, suggest that this chemical differentiation is intrinsic rather than a projection effect, reinforcing the importance of geometry when interpreting molecular tracers in protostellar environments.

\subsubsection{Streamers and warm spots}
Building on the clear identification of spiral-like streamers in Oph IRS 63, we detect analogous features in DK Cha, although with differing morphologies and S/Ns that likely reflect the influence of its more face-on orientation and varying shock activity (Fig. \ref{fig_moms_dkcha}).
Three spiral-like structures can be identified on the moment 0 map of SO (S1--S3; Fig. \ref{fig_moms_dkcha}). The stronger heads of S1 and S2 have S/Ns greater than 10, while the weaker tails of S1--S3 have S/Ns of approximately 5.
As observed in Oph IRS 63, we interpret these as accretion streamers of DK Cha.
It is evident that the south-east tail of S2 exhibits very narrow line widths (\mbox{$\Delta V \sim$ 0.5 km s$^{-1}$}; Fig. \ref{fig_dkcha_pv}).
The line width of S2 increases dramatically when it intersects with P1, a broad-line (FWHM $\sim$ 3 km s$^{-1}$) and SO-bright segment with a projected length of $\sim 300$ au.
P1 is also bright in SO$_2$ emission (panels d--f of Fig.~\ref{fig_moms_dkcha}).
The inner end of P1 is located at the boundary of the disk traced by the dust continuum.
There is no obvious velocity gradient along S2 and P1, which may be due to projection effects, given that DK Cha is nearly face-on and S2 lies along the major axis in the plane of the sky.
In addition to S2, a short streamer, S3, is also connected to P1.
The intersection between S3 and P1 produces a compact spot (with a size comparable to the beam), which is very broad in line width 
\mbox{(FWHM $\sim$ 4 km s$^{-1}$)}.
Below, this spot is denoted as $\rm S3\perp P1$.
We interpret that both S2 and S3 are falling onto the inner envelope: the falling gas forms the streamer P1, and P1 continuously feeds material to the central disk.
A warm spot is created when S3 hits P1.
P1 continuously endures shocks as it passes through the inner disk, leading to a long segment of emission of SO and SO$_2$ (Sect. \ref{sec_streamer_onoff}).

\begin{figure}[t]
\centering
\includegraphics[width=0.95\linewidth]{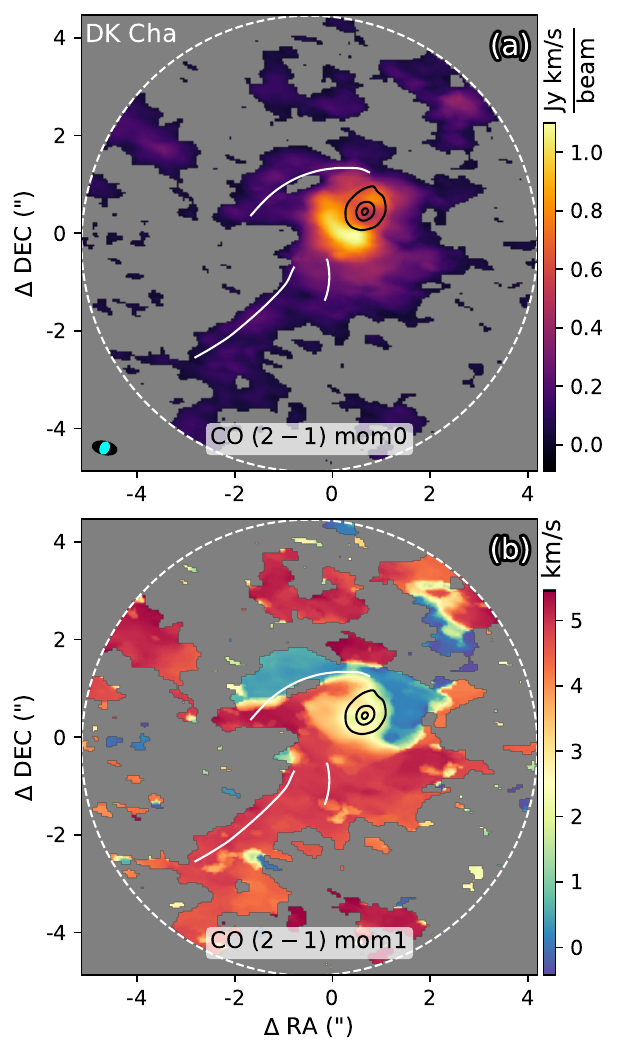}
\caption{Moment maps of CO (2-1) for DK Cha.
Only data within the velocity interval of 0–5 km s$^{-1}$ are utilized to calculate the moment maps (Sect. \ref{sec_dkcha_SandP}).
Velocity channels with intensities exceeding $3\sigma$ are included in the moment map calculation.
In each panel, data outside the white dashed circle have been masked out.
White curves represent the streamers identified from the SO emission (Fig. \ref{fig_moms_dkcha}).
The black contours denote the ALMA continuum, as shown in panel (a) of Fig. \ref{fig_moms_dkcha}.
The two ALMA observations were conducted in 2015 and 2023, respectively (Sect. \ref{sec_obs}). Position shifts due to proper motion and/or possible phase calibration uncertainties have been corrected by cross-matching their continuum peaks, determined via 2D Gaussian fitting.
In panel (a), the synthesis beams of the CO and SO data are represented by the black and cyan ellipses at the lower left corner, respectively.
\label{fig_dkcha_co}
}
\end{figure}

S1, the north streamer, is connected to a compact spot (denoted as P2 in the PV maps; Fig. \ref{fig_moms_dkcha}), which
has both emission of SO and SO$_2$.
P2 has a line width comparable to that of $\rm S3\perp P1$.
The excitation temperatures of SO$_2$ of the two spots
(P2, and $\rm S3\perp P1$) are calculated
through drawing the rotation diagrams of SO$_2$ (Fig. \ref{fig_dkcha_warmspots}).
P2 has a higher excitation temperature of SO$_2$ than
$\rm S3\perp P1$ (Sect. \ref{sec_hotspot}).
We interpret that S1 is a off-midplane streamer (Sect. \ref{sec_streamer_onoff}), and it directly hits the inner envelope near the centrifugal barrier, creating the hot spot P2.
The warm spots in DK Cha are generally hotter than those in Oph IRS 63, which is expected due to DK Cha's more massive central star.
Furthermore, DK Cha’s higher gas density, compared to Oph IRS 63, aids in detecting the high-$E_{\rm u}$ lines.

\subsubsection{A hot SO$_2$ spot of DK Cha?}\label{sec_hotspot}
Compared to the warm spot of $\rm S3\perp P1$, which has \mbox{$T_{\rm ex} \sim 70$ K}, the spot P2 exhibits an even higher temperature of \mbox{$T_{\rm ex} > 100$ K} (Fig. \ref{fig_dkcha_warmspots}).
Therefore, P2 can be considered a hot spot.
We examined the spectrum of SO$_2$ $16_{1,15}-15_{2,14}$, which has the highest $E_{\rm u}$ (131 K) among the observed lines toward DK Cha at P1 (Fig.~\ref{fig_dkcha_spw33spe}), and it shows a S/N higher than 5, confirming that the high fitted temperature is not caused by noise or spikes.
The temperature difference between the two spots can be explained by the differences in $\theta_{\rm view}$ and the specific angular momentum of the streamers that produce them.
S1 directly interacts with the midplane at P2, which is significantly closer to the center of mass compared to $\rm S3\perp P1$. Because S1 is off the midplane, it experiences less slowing down from the inner envelope before the strong shock occurs (Sect. \ref{sec_streamer_onoff}), leading to a hotter spot at P2.
The presence of multiple streamers interacting with the inner envelope and producing warm spot/segment—characterized by excitation temperatures notably higher than those in Oph IRS 63—highlights the influence of stellar mass and polar angle in governing the nature and detectability of shock signatures.

\subsubsection{Streamers traced from archived CO data}
\begin{figure}[t]
    \centering
    \includegraphics[width=0.8\linewidth]{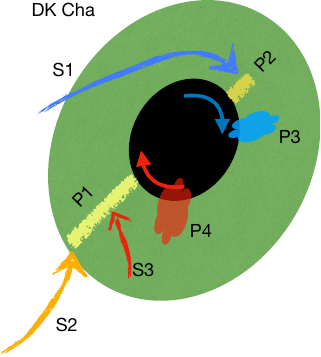}
    \caption{The schematic image of DK Cha based on its  emission of 
    SO and SO$_2$ (Fig. \ref{fig_moms_dkcha}).
    The major gas components that can be seen from the observed data (Fig. \ref{fig_moms_dkcha}) are labeled. The black ellipse marks the disk traced by
    continuum that cannot be fully resolved. The direction of rotation is unknown, 
    and we suggest a clockwise rotation based on the velocity distributions along the streamers and near the disk 
    traced by SO and CO (2-1) (Figs. \ref{fig_moms_dkcha} and \ref{fig_dkcha_co}; see also Sect. \ref{sec_dkcha_rotdirec}). The green ellipse is the assumed 
    disk-like inner envelope, which has strong interaction with the streamers that
    fall onto it (S1, S2, and S3). The interaction 
    contributes to the  enhancement of SO/SO$_2$ emission
    on the segment (P1) and spot (P2) at the inner envelope (Sect. \ref{sec_dkcha}).
    \label{fig_dkcha_cartoon}
}
\end{figure}

Consistent with the SO and SO$_2$ maps, archival CO (2-1) observations provide additional — albeit lower-resolution — confirmation of the streamer structures (Fig. \ref{fig_dkcha_co}). Two of the three streamers identified in SO (S1 and S2; Fig. \ref{fig_moms_dkcha}) are marginally visible in the moment 0 map of the archived CO (2-1) data. To reduce contamination from outflows, we integrated a narrow velocity interval (1–5 km s$^{-1}$) around DK Cha’s systemic velocity \citep[$\sim3$ km s$^{-1}$;][]{2012A&A...542A...8K}. The CO emission generally lies on the convex sides of the SO-traced streamers, suggesting outward forces acting upon them (Sect. \ref{sec_model}). Within the disk region traced by the ALMA continuum, CO (2-1) spectra show strong absorption near the systemic velocity. Moreover, the CO data have lower spatial resolution than the SO data (Sect. \ref{sec_obs}), so the moment 1 map of CO (panel (b) of Fig. \ref{fig_dkcha_co}) only roughly traces the disk’s velocity gradient (Sect. \ref{sec_dkcha_rotdirec}). A bright CO arc is visible at the southeastern disk margin, potentially marking the centrifugal barrier or the interaction between streamer/segment P1 and the inner envelope (Figs. \ref{fig_moms_dkcha} and \ref{fig_dkcha_cartoon}), although contamination from low-velocity outflow components cannot be excluded \citep{2023ApJ...945...63H}. Together, the CO data offer important complementary insights into the dynamics and external influences governing the infalling gas streams.

\subsubsection{Discussion of the direction of rotation}\label{sec_dkcha_rotdirec}
Extended SO emission features P3 and P4 (Fig. \ref{fig_moms_dkcha}) appear at blueshifted and redshifted velocities, respectively, but their origins are challenging to interpret (Fig. \ref{fig_dkcha_cartoon}). One possibility is that P4 represents the inner part of a streamer similar to S2 and S3, infalling onto the inner envelope rather than onto P1. P3 might simply be part of P2. Alternatively, these features could trace warm molecular gas influenced by outflows \citep{2006A&A...454L..75V}, or be shaped by the disk’s rotation.

Despite the wealth of spatial and spectral information, the exact rotational direction of DK Cha’s disk remains uncertain.
The morphology of S1 and the velocities observed in P1 and P2 suggest that DK Cha’s disk rotates clockwise (Fig.~\ref{fig_dkcha_cartoon}), a scenario supported by the velocity distribution of CO (2--1) within the disk region (Fig.~\ref{fig_dkcha_co}). However, this interpretation does not fully explain the morphology of S2. The radial velocity gradient of S2 (panel (b) of Fig.~\ref{fig_moms_dkcha}) differs from that expected in a simple rotational scenario, which may point to additional factors influencing its trajectory. One possibility—though speculative—is that S2 is affected by external pressure on larger scales (on the order of thousands of AU), perhaps from east to west, leading to reduced specific angular momentum. The angular momentum of S2 appears to decrease further as the gas enters P1, likely due to interactions between P1 and the inner envelope (Sects.~\ref{sec_shockedsurface} and \ref{sec_streamer_onoff}). Consequently, P1 moves inward toward the central mass with low specific angular momentum, and continuous shocks are expected as it propagates through the inner envelope. This scenario may explain the linear morphology and broad SO and SO$_2$ emission observed along P1. A schematic illustrating the connections between these streamers and features in DK Cha is provided in Fig.~\ref{fig_dkcha_cartoon}.

Previous observations have revealed multi-cone outflows in DK Cha with highly complex morphology \citep{2023ApJ...945...63H}. These intricate dynamical interactions, combined with external pressure and the multi-cone outflow structure, likely reflect the effects of variable angular momentum accretion and intermittent shock-driven phenomena, consistent with scenarios proposed for other protostellar systems \citep[e.g.,][]{2009ApJ...694.1045Z,2020A&A...636A..38N}.

\begin{figure}[t]
    \centering
    \includegraphics[width=0.99\linewidth]{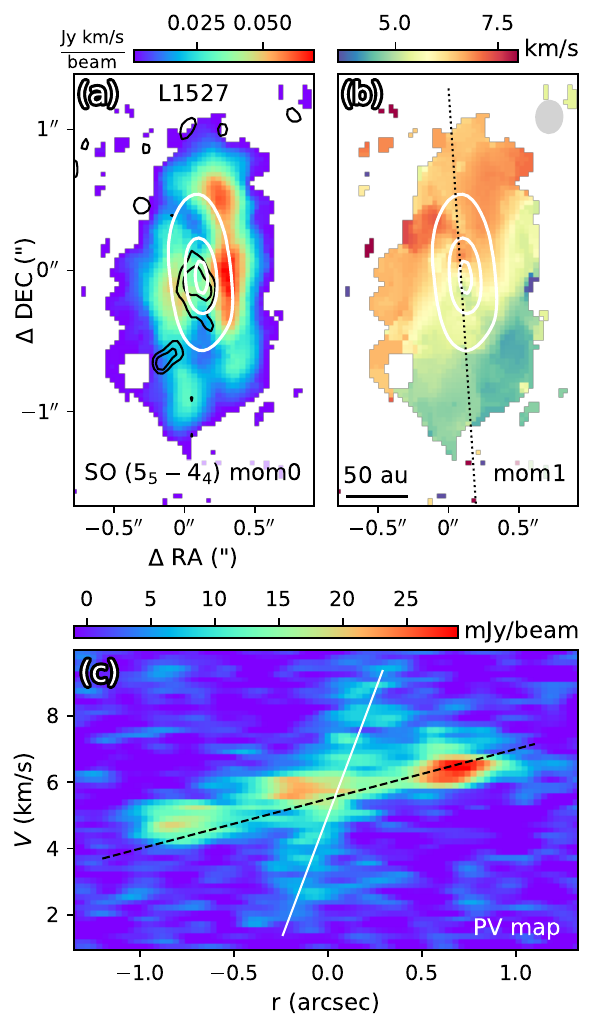}
    \caption{(a--b) Moment maps of SO of L1527. The white contours represent the continuum emission, with levels of 
    [0.1, 0.5, 0.9] times the peak value.
    The black contours represent the integrated intensity
    of SO$_2$ $16_{3,13}-16_{2,14}$, with levels of [8, 11]
    Jy beam$^{-1}$ km s$^{-1}$.
    The moment 0 map SO$_2$ is shown in Fig. \ref{fig_L1527_SO2}.
    In panel (b),
    the beam is shown at the upper-right corner by a gray ellipse.
    (c) PV map of SO along the black dotted line in panel (b). The white solid and black dashed 
    lines denote the two linear structures on the PV map.
    \label{fig_L1527_mompv}}
\end{figure}

\begin{figure*}[t]
    \centering
    \includegraphics[width=0.995\linewidth]{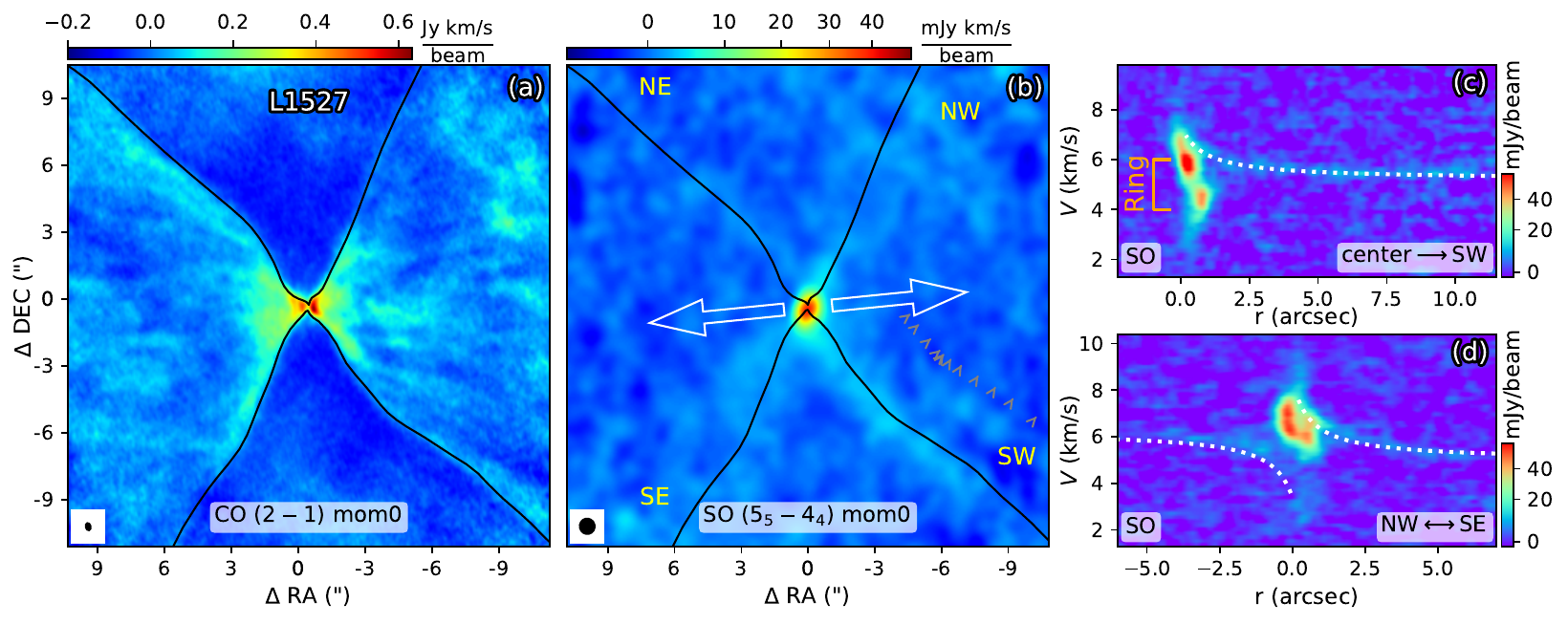}
    \caption{
    (a) Moment 0 map of CO (2-1) of L1527. The integration range is
    0.8--11.6 km s$^{-1}$. The beam size is shown as black ellipse at the lower left corner. The black lines mark the boundaries of 
    CO outflows.
    (b): Moment 0 map of SO, cleaned 
    using the {\it natural} weighting (Sect. \ref{sec_obssetup}).
    To improve the S/N, we have further smoothed the map to a beam size of 0.7\arcsec, as indicated by
    the black ellipse in the lower left corner. 
    The white arrows mark the outflows.
    The black lines have the same meaning as in the left panel,
    and are denoted as SW, SE, NW and NE, accordingly to their 
    locations. 
    (c) and (d): The PV maps of SO along the outflow boundaries 
    marked in the lower right corners. The dotted white lines
    represent the fitting results of  the trajectories of free-fall streams in PV space (Sect. \ref{sec_L1527_infalllayer}).
    See Figs. \ref{fig:lw_sw} and \ref{fig:lw_se_nw} for the typical spectra of SO along the outflow boundaries.
    \label{fig_L1527_cavity}
    }
\end{figure*}

\subsection{L1527} \label{sec_L1527}
In contrast to Oph IRS 63 and DK Cha, L1527, an edge-on Class 0 protostar, provides a different perspective on accretion shocks. The presence of accretion shocks in L1527, as traced by SO and SO$_2$ emission, offers valuable insights into how these features manifest in an edge-on system, shedding light on the dynamics of accretion in protostellar disks and envelopes.

\subsubsection{Parallel slabs of SO}
The SO emission from L1527 measured in this work (Fig. \ref{fig_L1527_mompv}) reveals two parallel slabs, nearly identical to those observed in different SO transitions in \citet{2014Natur.507...78S,2017MNRAS.467L..76S}. This supports the interpretation that SO emission arises from the surface of a disk or disk-like structure (see also Sects. \ref{sec_irs63_envelope} and \ref{sec_shockedsurface}). We only cover one SO$2$ transition for L1527 (Table \ref{tab_transitions}), which is marginally detected (Fig. \ref{fig_L1527_SO2}). Between the parallel slabs of SO, there is a compact region, approximately 50 au in size, that exhibits SO$2$ $16{3,13}-16{2,14}$ emission ($E_u = 148$ K), as shown by the black contours in Fig. \ref{fig_L1527_mompv} (see also Fig. \ref{fig_L1527_SO2}). Due to the limited data (only one transition), the temperature of the SO$_2$-emitting region cannot be well constrained. The SO$_2$ emission also appears to show two parallel slabs, but the low S/N prevents further analysis. Given that the SO$_2$ transition has a high excitation energy ($E_u \sim 150$ K, Table \ref{tab_transitions}), it likely traces a region that is warmer compared to the average temperature of the SO emission region.

\subsubsection{Two rings of SO at the disk scale}
\label{sec_L1527_2rings}


The PV map of SO along the major axis of L1527 (panel (c) of Fig.~\ref{fig_L1527_mompv}) reveals an X-shaped morphology composed of two linear strips, each showing a distinct velocity gradient.
Note that each linear strip corresponds to the projection of a rotating ring seen edge-on. This is because the projected velocity ($v_{\rm proj}$) and projected distance ($r_{\rm proj}$) vary with the sine of the azimuthal angle ($\phi$):
\begin{equation}
    v_{\rm proj} \propto r_{\rm proj} \propto \sin(\phi).
\end{equation}
The longer and brighter strip with the smaller velocity gradient corresponds to the outer ring (the parallel slab), which has been previously reported and interpreted as the centrifugal barrier of the envelope \citep{2014Natur.507...78S,2017MNRAS.467L..76S}.
In contrast, the fainter strip exhibits a much steeper velocity gradient (Fig.~\ref{fig_L1527_mompv}) and likely traces a newly identified inner ring of SO emission located inside the centrifugal barrier.
We refer to these features as the outer and inner rings, respectively.

The radii of the two SO rings inferred from the PV map are approximately $r_{\rm out} \sim 150$~au (or $\sim$1\arcsec) for the outer ring and $r_{\rm in} \sim 35$~au for the inner ring. The inner SO ring has a size comparable to the SO$_2$ emission region (Fig.~\ref{fig_L1527_mompv}). The corresponding velocity gradients observed on the PV map are $V_{\rm grad}^{\rm out} = 1.5$~km~s$^{-1}$~arcsec$^{-1}$ and $V_{\rm grad}^{\rm in} = 15$~km~s$^{-1}$~arcsec$^{-1}$, respectively. These values approximately obey the relation:
\begin{equation}
    \frac{V_{\rm grad}^{\rm out}}{V_{\rm grad}^{\rm in}} \sim \left( \frac{r_{\rm out}}{r_{\rm in}} \right)^{-1.5},
\end{equation}
suggesting that both rings are in Keplerian rotation. 
This follows because, for Keplerian motion, the rotational velocity satisfies
\begin{equation}
    v_{\rm rot} = \sqrt{\frac{GM_\star}{r}} \propto r^{-0.5}, \label{eq_keprot}
\end{equation}
and the corresponding velocity gradient scales as
\begin{equation}
    V_{\rm grad} \sim \frac{v_{\rm rot}}{r} \propto r^{-1.5}.
\end{equation}
Using Equation~\ref{eq_keprot}, the central mass of L1527 is estimated to be $\sim$0.5~$M_\odot$, consistent with the recent dynamical mass measurement (including the star and disk) by \citet{2023ApJ...951...10V} (see Sect.~\ref{sec_sample}).

A similarly steep linear structure was also observed in the PV map of H$_2$CO in L1527, corresponding to a radius of $\sim$25~au \citep{2023ApJ...951...10V}, which is consistent with the inner SO ring found in this work. That study interpreted the H$_2$CO structure as marking the snowline of H$_2$CO, located at a temperature of $\sim$70~K.
By analogy, the inner SO ring may represent the snowline of SO. The sublimation temperature of a molecular species depends on its binding energy, which remains uncertain \citep{2013A&A...550A..36M,2017ApJ...844...71P}. According to the UMIST database \citep{2013A&A...550A..36M}, the binding energy of SO is 2600~K, which is somewhat higher than that of H$_2$CO (2050~K), though other studies suggest a lower SO binding energy of $\sim$1800~K \citep{2017ApJ...844...71P}. This implies that the inner SO ring may indeed trace the SO snowline.
In this interpretation, SO is initially enhanced at the centrifugal barrier (outer ring), then depleted or destroyed within the outer disk, and re-enhanced near the snowline (inner ring). Similarly, SO$_2$ is also enhanced in the vicinity of the inner SO ring.

\begin{figure*}[t]
    \centering
    \includegraphics[width=0.99\linewidth]{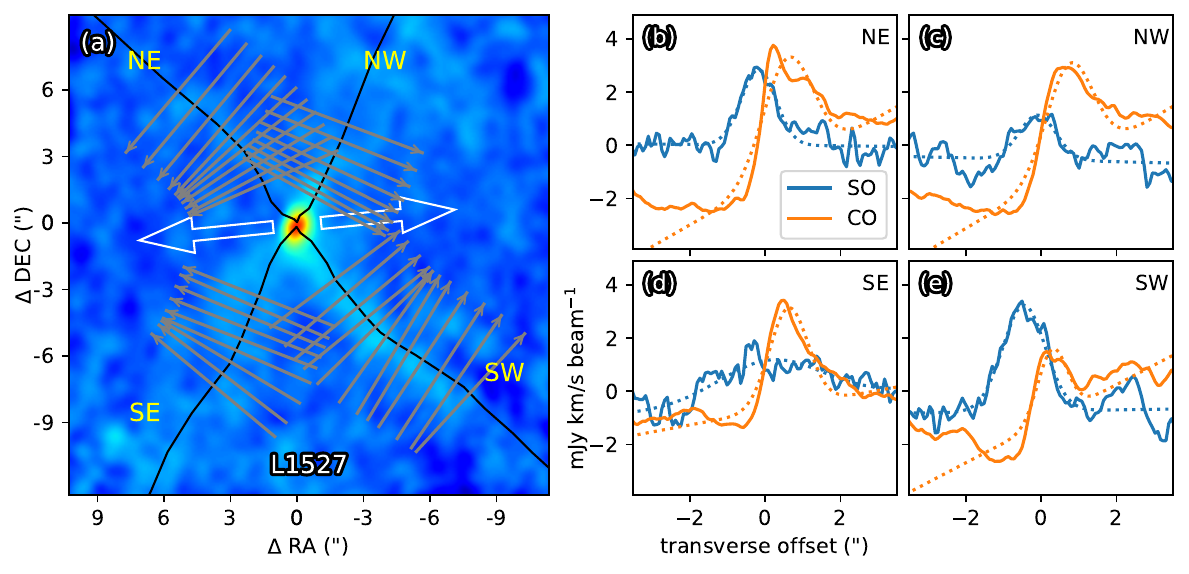}
    \caption{
(a) Same as Fig.~\ref{fig_L1527_cavity}, with gray arrows indicating the transverse directions along the outflow boundaries used for extracting radial profiles. The direction of each arrow denotes the positive offset direction, pointing toward the inner regions of the outflow cavities.  
(b–d) Averaged emission profiles along the transverse directions for the NE, NW, SE, and SW boundaries (corresponding to panel (a)).  
The SO emission is extracted from the image cleaned with natural weighting (without the additional smoothing applied in panel (a)).  
For comparison, the CO emission (orange) is divided by a factor of 30 and plotted alongside the SO emission (blue).  
Dotted lines show the corresponding fits using a Gaussian plus linear function.
}
    \label{fig_so_co_offset}
\end{figure*}

\subsubsection{Infalling layer of SO along the outflow boundary}
\label{sec_L1527_infalllayer}
We now turn to the large-scale SO emission.
No obvious streamers of SO are visible in the data cube cleaned using the {\it Briggs} {\it robust} = 0.5 weighting scheme (Sect. \ref{sec_obssetup}). 
Compared to this weighting scheme, the {\it natural} weighting ({\it robust} = 2) assigns equal weight to all samples, resulting in lower noise levels and a larger synthesized beam.  
Thus, the {\it natural} weighting enhances the S/N of large-scale emission.  
We examined the data cube of L1527, cleaned using {\it natural} weighting, to search for large-scale SO structures. Surprisingly, linear SO emission structures, referred to as SO strips, are visible along the boundaries of CO outflows (Fig. \ref{fig_L1527_cavity}). 
The most prominent SO strip lies along the southwestern (SW) outflow boundary, extending over 12\arcsec{} (1700 au) from the central protostar, 
with a S/N (at the emission peak) exceeding 5 in the moment 0 map.
The emission along the SW boundary is even more pronounced in the PV map, where the S/N exceeds 7. 
The SO spectrum stacked along the SW direction exhibits a very narrow linewidth of approximately 0.44 km s$^{-1}$ (see Appendix~\ref{sect_lw_sw} for details), confirming that the SO emission along the outflow boundaries is only partially resolved spectrally. 
Along the northwestern (NW) and southeastern (SE) outflow boundaries, SO emission is only marginally detected in the moment 0 map (panel (b) of Fig.~\ref{fig_L1527_cavity}). However, in the PV map (panel (d) of Fig.~\ref{fig_L1527_cavity}), narrow-line SO emission is clearly observed along these boundaries, with S/Ns of $\sim 5$.

A notable feature of the SO strips, particularly along the SW boundary, is that the SO emission peak is spatially offset from the CO emission peak, lying on the convex (outer) side of the outflow boundaries.  
To quantify this spatial offset between the two tracers, we present the averaged transverse emission profiles of SO and CO along the outflow boundaries in Fig.~\ref{fig_so_co_offset}. 
For the SW boundary, a significant offset is found between the Gaussian fits of the SO and CO profiles, with a $t$-statistic exceeding 15 (Table~\ref{tab_L1527_offset}), indicating a highly significant separation. 
The SO emission is shifted by approximately $0.7\arcsec \pm 0.01\arcsec$ relative to the CO emission in the transverse direction, suggesting that SO traces the convex side of the hourglass-shaped CO cavity walls. 
The FWHMs of the Gaussian fits are $1.3 \pm 0.1\arcsec$ for SO and $0.9 \pm 0.1\arcsec$ for CO. 
Note that the CO and SO images (with natural weighting) have similar beam sizes of $0.32\arcsec$ and $0.35\arcsec$, respectively. 
The narrow transverse profiles confirm the presence of well-defined strips in the moment 0 maps, with the CO boundary appearing sharper than the SO strip.
Similar spatial offsets between CO and SO are also observed along the NE and NW boundaries (Fig.~\ref{fig_so_co_offset}). 
Since the transverse directions differ for each boundary, these systematic offsets cannot be attributed to image misalignment between the CO and SO data.
This supports the interpretation that the SO emission originates from a conical layer that is either geometrically thicker or located farther from the outflow axis than the CO-emitting surface.

The SO spectra along the strips (outflow boundaries) exhibit central velocities close to the systemic velocity at large scales (6 km s$^{-1}$), with their velocity offset (relative to the systemic velocity) increasing significantly as the distance from the central protostar decreases (Fig.~\ref{fig_L1527_cavity}).  
This behavior suggests that the SO layer may be undergoing infalling motion rather than outflowing.  
To test this hypothesis, we model the strips in the PV maps using free-fall trajectories, following a similar approach to that in Fig.~\ref{fig_IRS63_model_streamers} (see details in Appendix \ref{sec_parabolic_trajectory}).  
In the model, we adopt a central mass of $M_\star = 0.5\ M_\odot$ (Sect.~\ref{sec_L1527_2rings}), a centrifugal barrier radius of $R_{\rm C} = 1\arcsec$, and viewing angles $\theta_{\rm view} = 90\degr$, $\phi_{\rm view} = 90\degr$, with initial trajectory angles $\theta_0 = 45\degr$ and $\phi_0 = 0\degr$.  
Remarkably, the free-fall trajectories reproduce the observed strips very well (panels (c) and (d) of Fig.~\ref{fig_L1527_cavity}).  
From the PV maps, it is evident that the southern strips along SE and SW do not directly connect to the southern blue part of the SO ring, but rather to its northern part.  
Similarly, the northern strip along NW appears connected to the southern blue part of the SO ring.  
This supports the interpretation that the SO strips do not move along straight paths but instead follow curved (possibly parabolic) trajectories wrapping around the outflow cone as they move inward.

Observations of protostars such as HH~211 and IRAS~16293$-$2422~A1 reveal that SO emission can exhibit velocity gradients perpendicular to the outflow axes, which have been interpreted as signatures of twisted magnetic fields \citep{2021ApJ...907L..41L,2021ApJ...921...12O}.  
Due to the limited S/N in our data, we are unable to place meaningful constraints on any velocity gradient perpendicular to the outflow axis in L1527.

Based on the preceding results, we propose the presence of a mildly shocked, conical SO-emitting layer undergoing downward-spiraling infall, which warps the outflow cavity.  
The observed SO emission pattern may result from interactions between the outflow and the infalling envelope, suggesting that angular momentum transfer from the outflow to the envelope is non-negligible (Sect.~\ref{sec_hardsurface}).  
We interpret the infalling SO layer as tracing the surface of the outer envelope on large scales (hundreds to over a thousand astronomical units).  
An increase in angular momentum leads to an outward expansion of the centrifugal radius.  
As a result, on smaller scales (from hundreds to several thousand au), the outflow boundary does not necessarily make direct contact with the inner envelope (Sect.~\ref{sec_hardsurface}).

\section{Shock confined outer and inner envelopes?}\label{sec_model}
In the previous sections, we have shown that SO and SO$_2$ emissions primarily trace shocks located near the centrifugal barrier and at the bases of infalling gas streamers. These observational features indicate that shocks play a significant role in shaping both the envelope structures and the disk-envelope interface. Furthermore, the spatial and kinematic patterns of SO emission near the outflow boundaries suggest possible interactions between the infalling envelope and the outflow. Although this evidence remains tentative, such interactions may contribute to compressing the outer envelope and consequently influence the motions of the inner envelope. The interaction between streams/streamers and envelope/disk serves as a key factor in the patterns of shock-related emission.  
Motivated by these observational insights, we constructed a self-consistent model of the disk-envelope system and the streamers to explore these possibilities further and to better understand the physical mechanisms driving the observed features.

\begin{figure*}[!thb]
\centering
\includegraphics[width=0.995\linewidth]{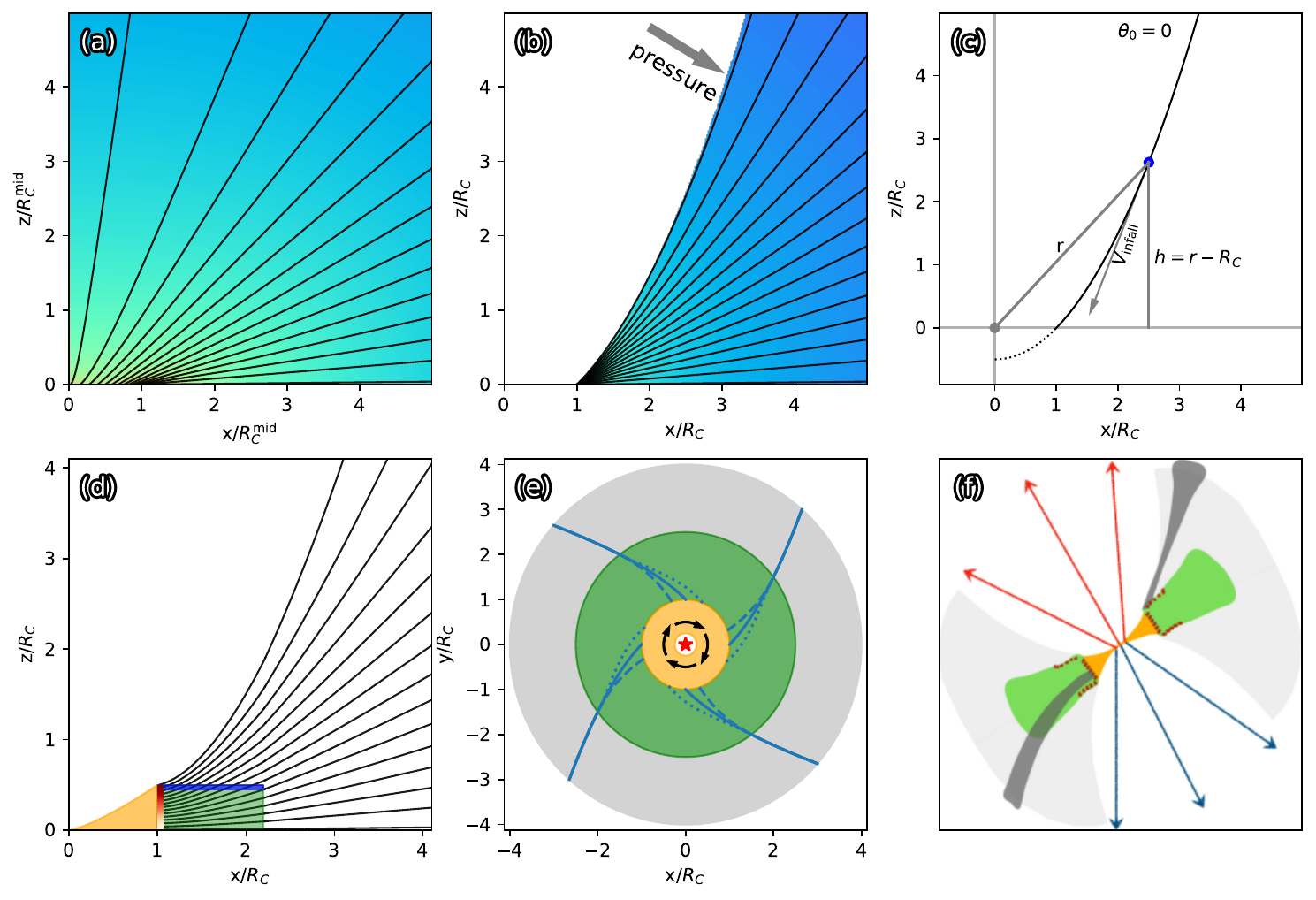}
\caption{(a) The density distribution of the \citet{1976ApJ...210..377U} model of the envelope. The black lines are the streamlines. 
(b) Same as the first panel, but for the constant-$J$ case (Sect. \ref{sec_hardsurface}).
(c) The streamline for $\theta_0 = 0$ in the constant-$J$ case. Note that $\rho$ depends only on $r$ in the colored region.
(d) The green area represents the disk-like envelope supported by thermal pressure and confined by shocks at the surface (Sect. \ref{sec_shockedsurface}). The orange area shows the schematic diagram of a Keplerian disk within the centrifugal barrier at $R_{\rm C}$, with a flare index of 1.3 \citep[e.g.,][]{1997ApJ...490..368C}. The vertical orange-red bar shows the head-on strong shocks at the centrifugal barrier (redder color means stronger shocks). The horizontal blue bar roughly represents the shocked surface of the envelope due to oblique flows.
(e) The streamlines in the envelope viewed from the pole. The gray and yellow regions represent the outer and inner envelopes, respectively. The blue solid line depicts the free-fall parabolic trajectory in the midplane. The dotted and dashed lines are the pressure-modified trajectories in the midplane and on the surface of the inner envelope, respectively.
(f) The schematic map in edge-on view. The gray, green, and orange regions are the outer envelope, disk-like envelope, and the disk, respectively. The black strips are the near-middle-plane (lower left) and off-midplane (upper right) streamers. The arrows indicate the outflows. The red dots highlight the shocked surfaces of the disk-like envelopes.
\label{fig_model_multipanel}
}
\end{figure*}

\begin{figure*}
    \centering
    \includegraphics[width=0.8\linewidth]{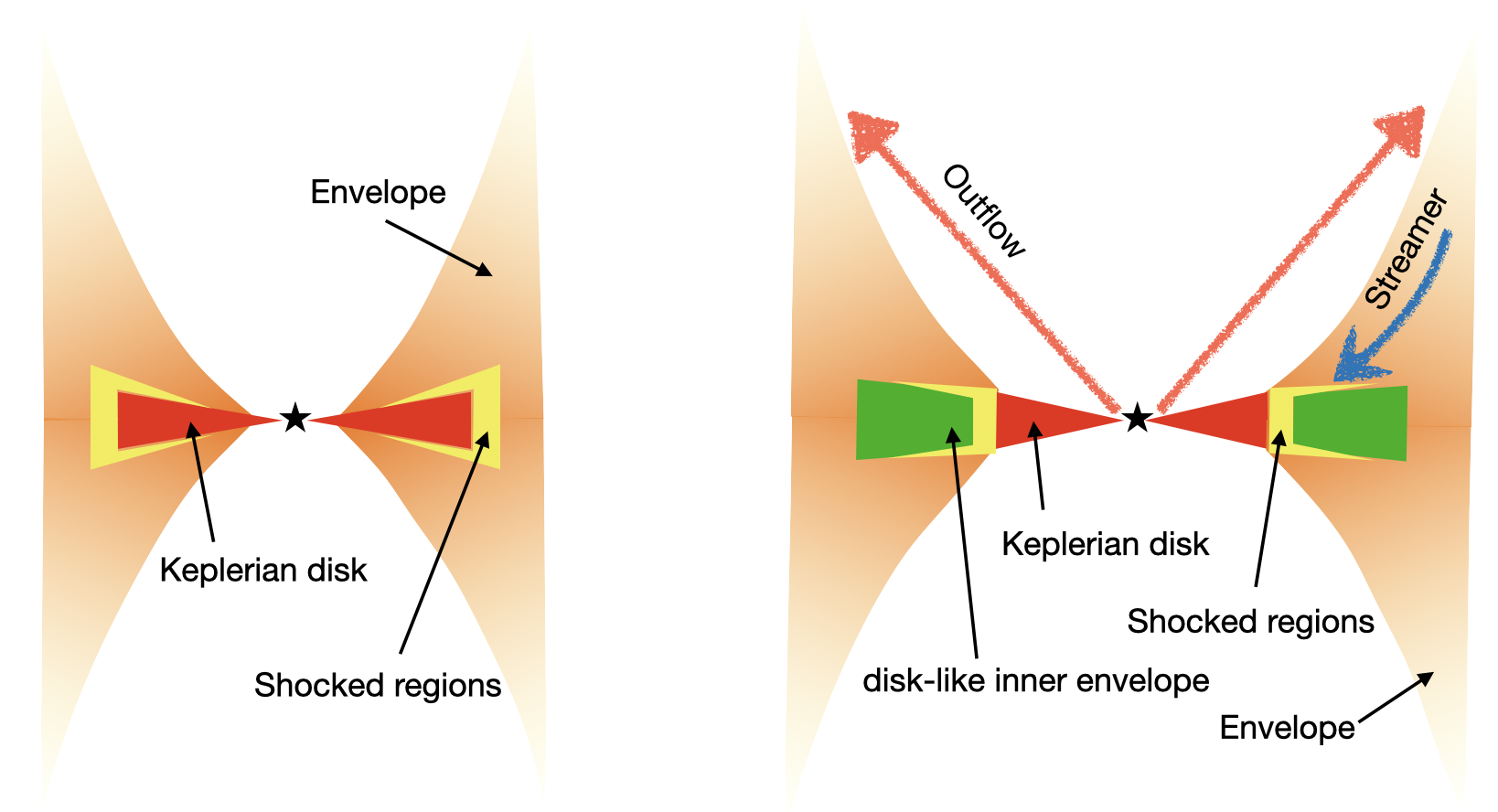}
    \caption{The schematic maps of the UCM model (left; see Sect. \ref{sec_cmu}) and the updated model presented in this work (right; see Sects. \ref{sec_hardsurface}, \ref{sec_shockedsurface}, and \ref{sec_streamer_onoff}). The right panel provides similar information to panel (f) of Fig. \ref{fig_model_multipanel}, but emphasizes a comparison between the locations of the shocked regions in the updated model and the UCM model.
     \label{fig_compare_models}}
\end{figure*}

\subsection{CMU model}\label{sec_cmu}

We begin by considering the classical ``CMU'' model of a protostellar envelope \citep{1976ApJ...210..377U,1981Icar...48..353C}. In this model, the gas moves in free fall with zero total energy (the sum of kinetic and gravitational potential energy). For a gas stream initially infalling with a polar angle $\theta_0$ (measured from the system’s rotation axis, which is typically defined by the angular momentum of the protostar and its surroundings), the specific angular momentum is given by
\begin{equation}
J = j(\theta_0) J_{\rm mid}.
\end{equation}
Here, $\theta_0$ is the initial polar angle of the gas trajectory relative to the rotation axis. By convention, $\theta_0 = 0$ corresponds to the rotation axis (pole), and $\theta_0 = \pi/2$ corresponds to the midplane (equator). At the midplane, $j(\pi/2) = 1$, meaning the specific angular momentum is largest there.

Ignoring gas pressure, the gas stream follows a parabolic trajectory. The radius of the centrifugal barrier—defined as the distance from the focus to where the parabolic path crosses the midplane—is
\begin{equation}
R_{\rm C}(\theta_0) = \frac{J^2}{GM_\star} = j^2(\theta_0) R_{\rm C}^{\rm mid},
\label{eq_rc}
\end{equation}
where
\begin{equation}
R_{\rm C}^{\rm mid} = \frac{J_{\rm mid}^2}{GM_\star}.
\end{equation}
This $R_{\rm C}^{\rm mid}$ is the centrifugal barrier radius for gas falling exactly in the midplane.

Under the gravity of the central star (ignoring the envelope’s self-gravity), the parabolic trajectory is described by \citep{1976ApJ...210..377U}
\begin{equation}
\frac{r}{R_{\rm C}^{\rm mid}} = \frac{j^2(\theta_0) \cos(\theta_0)}{\cos(\theta_0) - \cos(\theta)},
\label{eq_sl}
\end{equation}
where $r$ and $\theta$ are the radius and polar angle at a given point along the streamline.

The radial infall velocity (velocity component along $r$) is \citep{1981Icar...48..353C}
\begin{equation}
v_r = -\left(\frac{GM_\star}{r}\right)^{1/2} \left(1 + \frac{\cos(\theta)}{\cos(\theta_0)}\right)^{1/2}.
\label{eq_velo_r}
\end{equation}
Although $v_r$ depends on $r$, $\theta_0$, and $\theta$, substituting Equation~\ref{eq_sl} removes the explicit dependence on $\theta$.

Assuming the accretion rate at large distances is spherically symmetric, mass conservation implies the gas density can be expressed as \citep{1981Icar...48..353C}
\begin{equation}
\begin{aligned}
\rho &= \frac{\dot{M}_{\rm tot}}{4\pi r^2 v_r} \left(\frac{\partial \theta}{\partial \theta_0}\right)^{-1} \frac{\sin(\theta_0)}{\sin(\theta)} \\
&= \frac{\dot{M}_{\rm tot}}{4\pi r^2 v_r} \left( \left(1 - \frac{R_{\rm C}}{r}\right) + \frac{\cos(\theta_0)}{\sin(\theta_0)} \frac{d R_{\rm C}}{r d \theta_0} \right)^{-1}.
\label{eq_rho_raw}
\end{aligned}
\end{equation}
The original CMU model assumes
\begin{equation}
j(\theta_0) = \sin(\theta_0),
\end{equation}
which leads to
\begin{equation}
\rho = \frac{\dot{M}_{\rm tot}}{4\pi r^2 v_r} \left(1 + \frac{R_{\rm C}^{\rm mid}}{r} \left(2 - 3 \sin^2(\theta_0)\right)\right)^{-1}.
\label{eq_rho_cmu}
\end{equation}
The density depends on both radius $r$ and polar angle $\theta$ (see panel (a) of Fig.~\ref{fig_model_multipanel}). 
Taking $R_{\rm C}^{\rm mid}$ as the disk radius, the mass infall rate from the envelope onto the disk within radius $R < R_{\rm C}^{\rm mid}$ is
\begin{equation} \label{eq_Mdot_cmu}
\dot{M}(R) = \left(1 - \sqrt{1 - \frac{R}{R_{\rm C}^{\rm mid}}}\right) \dot{M}_{\rm tot}.
\end{equation}

Eq. \ref{eq_Mdot_cmu} leads to that half of the envelope’s mass falls onto the disk within $0.75\, R_{\rm C}^{\rm mid}$. 
However, this conflicts with observations showing that shock tracers such as SO and SO$_2$ are lacking in the central disk region (Sect.~\ref{sec_obsresults}). One possible explanation is that stellar radiation destroys these molecules during infall. But since the SO/SO$_2$ snowline is expected to lie very close to the star in low-mass protostars (see Sect.~\ref{sec_L1527} and references therein), radiation may not fully destroy these species.
A more likely explanation is that some mechanism prevents envelope gas from directly falling onto the inner disk regions.

\subsection{constant-$J$ case}\label{sec_warpmodel}
The model of \citet{1976ApJ...210..377U} requires a gas stream with a small $\theta_0$ (initially from near the pole) to have very little angular momentum. This may not hold under the influence of stellar feedback. Outflows and jets are common in protostars and are believed to clear the polar region, forming a cavity \citep[e.g.,][]{2016ARA&A..54..491B,2020ApJ...900...15L}. The cavity would exert continuous pressure on the gas stream originating near the pole during the early stages of accretion, efficiently increasing the stream's angular momentum. This is supported by the infalling layers of SO closely tracing the outer edges of the CO outflow boundaries (see example of L1527 in Sect.~\ref{sec_L1527_infalllayer}). Thus, $j(\theta_0)$ is not necessarily required to be an increasing function of $\theta_0$. Note that a stream with a larger $j(\theta_0)$ will have a larger $R_{\rm C}$ (Eq.~\ref{eq_rc}). If $j(\theta_0)$ is a decreasing function of $\theta_0$, the streamlines from different $\theta_0$ values will interact with each other before reaching their centrifugal barriers. However, considering that the inner pressure of the outflow could influence the system at a large scale (e.g., thousands of au), there is enough time for the pressure gradient within the envelope to adjust the angular momentum of different streams. Therefore, we adopt a constant-$J$ case with
\begin{equation}
j(\theta_0) = 1,
\label{eq_constj}
\end{equation}
and denote it as the constant-$J$ model.

\subsubsection{Hard-surface envelope by stellar feedback} \label{sec_hardsurface}
In the constant-$J$ model, the centrifugal barrier radius $R_{\rm C}$ is constant and does not depend on the initial polar angle $\theta_0$. Ignoring pressure effects, gas streams originating from different $\theta_0$ values will converge and interact at the same radius $R_{\rm C}$ (Eq.~\ref{eq_rc}). The density of the infalling gas can then be expressed as
\begin{equation}
\rho = \frac{\dot{M}}{4\pi r^2 v_r} \frac{\cos{\theta_0}}{\cos{\theta}}, \label{eq_rhoenv}
\end{equation}
and using Eqs.~\ref{eq_sl} and \ref{eq_constj}, one obtains the relation
\begin{equation}\label{eq_theta_r}
\frac{\cos(\theta)}{\cos(\theta_0)} = \frac{r - R_{\rm C}}{r}.
\end{equation}
Substituting Eq.~\ref{eq_theta_r} into Eq.~\ref{eq_rhoenv} gives a simplified expression for the density:
\begin{equation}
\rho = \frac{\dot{M}}{4\pi r^2 v_r} \frac{r}{r - R_{\rm C}}. \label{eq_rhoenv_final}
\end{equation}
This is equivalent to Eq.~\ref{eq_rho_raw} when a constant $R_{\rm C}$ is assumed. Note that in the constant-$J$ model, $\rho$ explicitly depends only on $r$ and not on $\theta$ or $\theta_0$. Physically, this implies that at a given radius $r$, the infalling gas is uniformly compressed in the polar angle ($\theta$) direction for $\theta > \theta(r)|_{\theta_0=0}$, while the density sharply drops to zero for $\theta < \theta(r)|_{\theta_0=0}$. This discontinuity defines a "hard surface" for the envelope (panel (b) of Fig.~\ref{fig_model_multipanel}). Note that external pressure sources, such as outflows, may only become significant at large radii. At smaller radii, the envelope surface does not need to be in direct contact with or confined by stellar feedback. Recent near-infrared JWST observations support the presence of an empty region between the outflow cavity and the inner envelope surface \citep[e.g.,][]{2023ApJ...951L..32H}.

Similarly, substituting Eq.~\ref{eq_theta_r} into Eq.~\ref{eq_velo_r} yields
\begin{equation}
v_r = -\left(\frac{GM_\star}{r}\right)^{1/2} \left( \frac{2r - R_{\rm C}}{r} \right)^{1/2}.
\end{equation}
Thus, the radial infall velocity $v_r$ also becomes independent of both $\theta$ and $\theta_0$. Therefore, in the constant-$J$ model, both $v_r$ and $\rho$ explicitly depend only on the radius $r$ (see panel (b) of Fig.~\ref{fig_model_multipanel}). For a stream lying in the midplane, the azimuthal velocity is given by
\begin{equation}
v_{\phi} = \left(\frac{2GM_\star}{r} - v_r^2\right)^{1/2} = \sqrt{GM_\star R_{\rm C}} \, r^{-1}, \label{eq_vphi}
\end{equation}
where $v_r$ is given by Eqs.~\ref{eq_velo_r} and \ref{eq_theta_r}.

\subsubsection{Disk-like inner envelope confined by accretion 
 shocks}\label{sec_shockedsurface}
The model above considers only the streams in free fall. As $r$ approaches the centrifugal barrier, the density increases drastically. When the streams reach the centrifugal barrier, shocks occur. A pressure gradient will develop due to the accumulation of material and the rise in temperature. As a result, the inner region of the envelope will swell (panel (d) of Fig.~\ref{fig_model_multipanel}).

For the stream with $\theta_0 = 0$ (initially from the pole), when it enters the swell-up region, it will experience two types of decelerating forces: one from the ram pressure and the other from the thermal pressure gradient. The ram pressure force is aligned with the direction of gas motion ($-V_{\rm infall}$). The thermal pressure gradient, on the other hand, is directed along $r$, since $\rho$ is constant as a function of $\theta$ at a fixed $r$ (see Sect.~\ref{sec_hardsurface}). We denote $h = z(r, \theta) = r \cos(\theta)$ as the height of the stream above the midplane, and $v_h$ as the velocity of the stream approaching the midplane. According to the geometric properties of a parabola (panel (c) of Fig.~\ref{fig_model_multipanel}), at a given point on the streamline ($r, \theta$),
\begin{equation}
v_r \equiv v_h
\end{equation}
and
\begin{equation}
h = r - R_{\rm C}.
\end{equation}
The ram pressure will decelerate both $v_r$ and $v_h$ equally, allowing the stream to continue moving toward the centrifugal barrier. In contrast, the thermal pressure gradient exerts a stronger decelerating effect on $v_r$ than on $v_h$, causing the stream to fall onto the midplane before reaching the centrifugal barrier. Shocks will occur between the stream and the swell-up region if the residual of $v_h$ (denoted as $v_h^{\rm diff}$),
\begin{equation}
v_h^{\rm diff} = \max(v_h - v_r),
\end{equation}
exceeds the sound speed.

Confined in the $h$ direction by shocks at its surface, the swell-up region will then be shaped into a flattened disk-like envelope (as qualitatively represented by the green area in panel (d) of Fig.~\ref{fig_model_multipanel}). The detailed distribution of the velocity of the disk-like inner envelope is complex. For simplicity, we propose the following for a stream in the midplane:
\begin{align}
   & v_\phi \propto r^{-1}, \label{eq_mod_vphi}\\
   & v_r \propto r^0 \propto \text{const.} \label{eq_mod_vr}
\end{align}
That is, $v_r$ is constrained to a constant value (with a power-law index $\beta = 0$), while $v_\phi$ remains largely unaffected compared to Eq.~\ref{eq_vphi}. This results in a rotating-and-infalling disk-like inner envelope, with a steeper decrease in $v_\phi$ ($\beta = -1$) as $r$ increases, compared to Keplerian motion ($\beta = -1/2$). The surface of the disk-like inner envelope, where shocks occur, experiences shear forces from the gas beneath it. Consequently, the exact value of $\beta$ is challenging to predict, falling within the range of 0 to $-1$. If the motion of the post-shock surface is entirely governed by the gas below it, then Eqs.~\ref{eq_mod_vphi} and \ref{eq_mod_vr} provide a reasonable description of the velocity distribution at the post-shock surface. This scenario is supported by our observations (Figs.~\ref{fig_irs63_SO_mom12} and \ref{fig_pv_irs63}).

The proposed shock-confined disk-like envelope shares some similarities with the so-called `pseudodisk' supported by the $B$ field \citep[e.g.,][]{2016ApJ...822...12H}. However, there are significant differences. We do not incorporate the $B$ field in our model, as recent simulations \citep[e.g.,][]{2019MNRAS.489.1719W} suggest that magnetic braking may not be crucial for the formation of low-mass disks. The disk-like envelope is located at $R > R_{\rm C}$, rather than wrapping around the disk from all sides. We highlight the role of shocks at the surface of the disk-like envelope in both the formation of the inner disk-like envelope and the associated shock-related chemistry. The shock tracers will appear as a ring at a radius of $\sim R_{\rm C}$ in the face-on view, and as a parallel slab in the edge-on view, consistent with observations (Sect.~\ref{sec_obsresults}).

\subsubsection{Streamers dropping within/onto the disk-like envelope}
\label{sec_streamer_onoff}
In recent years, increasing observations have suggested that gas streamers, which connect the envelope to the disk, exist in protostars at various stages of development \citep{2020NatAs...4.1158P,2022A&A...667A..12V}. \citet{2020NatAs...4.1158P} first confirmed that these streamers can be explained as freely infalling gas toward the protostar. Consequently, it is natural to interpret streamers as a special class of streams that coexist with the globally infalling envelope. Streamers may also be linked to larger-scale filaments outside the envelope \citep[e.g.,][]{2023A&A...677A..92V} and to smaller-scale spiral-like structures on the disk \citep[e.g.,][]{2019Sci...366...90A}. A free-fall streamer and the projection of its parabolic trajectory can be modeled as described in Sect.~\ref{sec_parabolic_trajectory}. However, when an in-midplane streamer interacts with the disk-like envelope, its infall velocity is decelerated, resulting in a spiral-like structure with a large $\Delta \phi$ (see definition in Sect.~\ref{sec_hardsurface}). The slim morphology and much higher density of streamers in the inner few hundred au, compared to the global envelope, reflect the non-isotropic accretion characteristics of a young star as it interacts with its natal environment \citep[e.g.,][]{2020A&A...635A..67H}.

A streamer does not necessarily have the same specific angular momentum as the streams in the global envelope. Based on the different patterns of interaction between the streamers and the global envelope, we divide the streamers into two types: the in-midplane streamer and the off-midplane streamer (see panel (f) of Fig.~\ref{fig_model_multipanel} for illustration). Before it hits the centrifugal barrier, the in-midplane streamer will experience continuous interaction with the disk-like envelope. Relatively slow shocks occur during the descent of the streamer within the inner envelope, resulting in a shock-endured slim segment. Shock tracers such as SO will be enhanced and excited on this shock-enduring segment (e.g., the P1 of DK Cha shown in Fig.~\ref{fig_moms_dkcha}). This process increases the angle of incidence, thereby reducing the collision velocity when the streamer strikes the centrifugal barrier. The warm spot of SO$_2$, which requires stronger shocks than SO for excitation, will not necessarily appear at the foot of an in-midplane streamer.

In contrast, the off-midplane streamer will freely hit the surface of the disk or disk-like envelope near the centrifugal barrier. The angle of incidence tends to be small (panels (e--f) of Fig.~\ref{fig_model_multipanel}), corresponding to a nearly head-on collision. There will be no SO-enhanced segment. Instead, a warm and compact spot of SO$_2$ is expected at its foot (see Sects.~\ref{sec_warmspot} and \ref{sec_hotspot} for examples).

Overall, the constant-$J$ model offers a useful framework that highlights the key role of angular momentum in driving gas stream convergence and shock formation at/around the centrifugal barrier, leading to the formation of a flattened, disk-like inner envelope. The interactions between infalling streams and this structure influence mass transport and chemical processes around the disks of protostellar systems. This model provides a coherent interpretation consistent with our observations, helping to illuminate the complex dynamics of protostellar environments.

\section{Discussion}\label{sec_discussion}

\subsection{Links between observations and models}

The preceding sections have explored connections between models and observations; here, we examine this relationship in more detail. The introduction of a disk-like inner envelope is motivated by several considerations: 
(1) SO emission within the disk, inside the centrifugal ring, is very weak, whereas a weak but extended SO component is observed beyond the centrifugal ring;  
(2) Some streamers undergo significant deceleration before reaching the centrifugal barrier;  
(3) Certain streamers impact the midplane, producing warm spots outside the centrifugal barrier;  
(4) Long segments exhibiting broad SO and low-energy SO$_2$ emission exist beyond the centrifugal barrier, indicative of continuous shocks;  
(5) Outflows, common in protostellar systems, make it unlikely for gas flowing from the pole with very low angular momentum to directly collide with the inner disk;  
(6) An infalling SO layer may have been observed along the outflow boundary.

The original CMU model (Sect.~\ref{sec_cmu}) proposed accretion shocks at the centrifugal barrier and on the disk surface inside it. In contrast, our model (Sect.~\ref{sec_warpmodel}) suggests that shocks also occur at the surface of a disk-like inner envelope beyond the centrifugal barrier (Fig.~\ref{fig_compare_models}). The actual scenario likely lies between these models, with some material flowing inward above the inner disk surface. Our model highlights the important role of shocks in structure formation at the disk/envelope scales, complementing scenarios observed at both larger scales \citep[e.g., cloud scale;][]{1998ApJ...504..223G,2025arXiv250210897L,2025arXiv250220458L} and smaller scales \citep[inner disk scale;][]{2022ApJ...934L..20B}. Although the specific manifestations may differ (e.g., accretion, outflow, turbulence), shocks are fundamental in shaping astrophysical environments across a wide range of scales.

Besides helping shape the disk-envelope structure, accretion shocks could also play a key role in resetting the chemistry of gas accreting onto the disk (Sect.~\ref{sec_intro}). The shock location—whether inside or outside the centrifugal barrier—does not fundamentally alter this chemical “reset” concept, but it may affect the composition of the post-shock gas. Gas accreting onto the inner envelope likely experiences shocks under relatively weak UV radiation, while accretion shocks and stellar feedback (e.g., heating, UV radiation) may occur at separate stages. This is supported by the two rings seen in L1527’s SO PV map (Sect.~\ref{sec_L1527}). In contrast, gas falling directly onto the disk may encounter shocks and stellar feedback simultaneously, potentially leading to different chemical outcomes \citep[e.g.,][]{2021A&A...653A.159V,2023A&A...675A..86K}. Importantly, under the CMU model, most material undergoes strong shocks before entering the disk, whereas in our model, a significant fraction flows through the inner envelope and may avoid strong shocks. Observations of the edge-on source L1527 support this, with strong SO emission confined to disk surfaces \citep{2014Natur.507...78S,2017MNRAS.467L..76S}. This is consistent with simulations showing that much of the envelope material may accrete smoothly without strong discontinuities \citep[e.g.,][]{2024A&A...686A.253M}. Thus, material entering the disk via the inner envelope may be a mixture of both “modified” and “reset” gas (Sect.~\ref{sec_intro}). While a detailed chemical study is beyond the scope of this work, these findings provide a foundation for future investigations into how much the chemistry of infalling gas is reset before incorporation into planet-forming disks.

\subsection{Beyond the observations and models}\label{sec_beyonddis}

Observationally, this work focuses on the disk-envelope scale. Gas in the inner envelope shows significant infall motion (Sect.~\ref{sec_irs63_envelope}), resulting in a surface density much lower than that of the disk traced by millimeter continuum emission dominated by Keplerian rotation. This likely explains why the millimeter continuum size approximately coincides with the inner boundary of the centrifugal barrier traced by the SO ring. While CO (2--1) and SO lines effectively trace emission near the outflow boundaries, probing the inner envelope regions beneath the surface remains challenging. Observations of lower-$J$ transitions, such as CO ($J=1$--$0$) and N$_2$H$^+$, may better constrain these inner layers \citep[e.g.,][]{2021ApJ...908..108F}, motivating future studies to focus on such tracers.

As discussed in Sect.~\ref{sec_streamer_onoff}, streamers can exhibit specific angular momenta that differ significantly from the global envelope, with some possibly accreting directly onto the disk inside the centrifugal barrier. Due to limited spatial resolution and sample size, this work primarily focuses on shock-related SO and SO$_2$ emission at and beyond the centrifugal barrier. In Oph IRS 63, the SO moment 2 map reveals a spiral-like, broad-line ‘S’ shape (Fig.~\ref{fig_irs63_Smark}), which lacks a clear counterpart in the moment 0 map (Fig.~\ref{fig_irs63_color_maps}). The PV diagram along this ‘S’ structure (Fig.~\ref{fig_irs63_pvmaps_ringandS}a) shows two velocity components in the southeast segment, consistent with shocks confined to the inner envelope surface (Sect.~\ref{sec_obsresults}). Given the similarity in clockwise spiraling between this ‘S’ shape and the streamers, this feature may trace accretion flows funneling material from the inner envelope onto the disk \citep[e.g.,][]{2016A&A...590A..22H,2019Sci...366...90A,2020NatAs...4..142L}.

Turning to the modeling aspect, this study does not include the effects of magnetic fields on streamer and envelope dynamics \citep[e.g.,][]{2016ApJ...822...12H,2019MNRAS.489.1719W,2020MNRAS.491.2180M,2023ASPC..534..317T,2023ASPC..534..465L}, nor does it explore the detailed nature of shock types \citep{2015A&A...578A..63F}. These factors could influence interpretations, especially given the current lack of B-field tracer data. Additionally, the sources examined here are primarily low- to intermediate-mass protostars. For massive young stellar objects, stronger stellar feedback and clustered environments \citep[e.g.,][]{2019A&A...632A..57C,2023ASPC..534..275O,2024RAA....24b5009L} may expose infalling material to intense radiation before it reaches the centrifugal barrier, altering its dynamics and chemistry.

Despite these limitations, the good agreement between observations and our descriptive model suggests that the accretion shock framework remains a valid approach for understanding the disk-envelope interface in low- and intermediate-mass protostars. Moving forward, expanding observations to larger and more diverse samples—utilizing a variety of shock tracers and targeting protostars with different masses and inclination angles—will be essential for fully characterizing the physical and chemical processes associated with accretion shocks.

\section{Summary}\label{sec_summary}

In this work, we searched for and explained signatures of accretion shocks by examining the spectral line emission of SO and SO$_2$ observed by ALMA toward several nearby protostars at different inclination angles ($\theta_{\rm view}$). Streamers are commonly observed in these sources (Oph IRS 63 and DK Cha). SO emission tends to concentrate at the centrifugal barrier of the envelope, forming a ring-like structure (Oph IRS 63 and L1527). Outside the centrifugal barrier, SO emission is widely distributed or associated with streamers interacting with the envelopes. Warm spots or segments of SO$_2$ were detected at the feet of the streamers. We proposed a disk-like inner envelope model to consistently explain the locations and mechanisms of SO/SO$_2$-related accretion shocks. Below, we summarize the key conclusions of this work:

\begin{itemize}
\item[1.] 
In Oph IRS 63 ($\theta_{\rm view} \sim 45\degr$), we found an SO$_2$ ring associated with the SO ring surrounding the inner protoplanetary disk. The SO/SO$_2$ ring traces the centrifugal barrier of the infalling envelope at a radius of $\sim 80$ au. 
Outside the SO/SO$_2$ rings, the velocity distribution of the weak and extended SO emission can be explained by a rotating-and-infalling inner envelope. Four spiral-like streamers are visible in the SO emission (denoted as S1 to S4). A warm spot of SO$_2$ (with $T_{\rm ex} \gtrsim 50$ K) is found at the foot of S3, and we suggest that S3 is an off-midplane streamer that violently hits the centrifugal barrier.

\item[2.] 
In DK Cha ($\theta_{\rm view} \sim 20\degr$), a long segment of $\sim 300$ au with broad-line (FWHM $\sim$ 4 km s$^{-1}$) emission of SO and SO$_2$ is observed. Two streamers are falling onto the segment, and we speculate that the interaction between the segment and the inner envelope is responsible for the enhancement of SO/SO$_2$. 
The segment and another streamer feed material into the central disk. A warm spot ($\sim 70$ K) and a possible hot spot ($\gtrsim 100$ K) of SO$_2$ are found at the feet of the streamers. The streamers are also visible in CO emission, located along the outer edges of the SO emission.

\item[3.] 
In L1527 ($\theta_{\rm view} \sim 90\degr$), the SO emission is consistent with prior observations \citep{2014Natur.507...78S,2017MNRAS.467L..76S}, showing the shape of parallel slabs. This supports the idea that SO emission occurs in ring-like patterns at the surface of a disk or disk-like structure.  
An `X' shape is seen in the SO PV map, which we interpret as two rings. The velocities of the two rings can be explained by Keplerian motion around a central mass of $\sim$0.5 $M_\sun$. The outer ring corresponds to the centrifugal barrier, as noted in the literature. The inner ring may trace the SO snowline.  
We also detect SO strips along the outer edges of the CO outflow boundaries, which can be reasonably modeled by free-fall trajectories, indicating likely interaction between the outflows and the large-scale envelope surfaces. We propose that there may be an infalling layer warping the outflow cavity, which endures mild shocks.

\item[4.] 
The model proposed in this work suggests the presence of a disk-like inner envelope outside the centrifugal barrier, formed by the compression of constant-$J$ streams that partly acquire their angular momentum from outflows. While the centrifugal barrier follows Keplerian motion, the disk-like inner envelope exhibits both rotation and infall. Shocks at its surface contribute to the weak, extended SO emission observed beyond the centrifugal barrier. Streamers interacting with this inner envelope deviate from free-fall trajectories, enhancing SO and SO$_2$ emission in their inner regions. Midplane streamers experience continuous shocks, forming warm segments, whereas off-midplane streamers tend to create warm or hot spots due to more violent collisions. The model is consistent with various observational evidences presented in this work (e.g., sulfur ring, SO streamers, warm spots/segments, envelope motion, infalling layers around outflows).
\end{itemize}

Overall, SO and SO$_2$ are excellent tracers of accretion shocks around disks. The shock-related chemistry at the surfaces of the disk, the centrifugal barrier ring, and the disk-like inner envelope warrants special attention when studying the chemical and physical properties of protoplanetary disks.

\begin{acknowledgement}
X.L. acknowledges the support of the Strategic Priority Research Program of the Chinese Academy of Sciences  under Grant No. XDB0800303,
and the National Key R\&D Program of China under Grant No. 2022YFA1603100.
This paper makes use of the following ALMA data: ADS/JAO.ALMA\#2022.1.01411.S,
\#2023.1.00592.S,
\#2013.1.00708.S,
\#2019.A.00034.S. 
ALMA is a partnership of ESO (representing its member states), NSF (USA) and NINS (Japan), together with NRC (Canada), NSTC and ASIAA (Taiwan), and KASI (Republic of Korea), in cooperation with the Republic of Chile. The Joint ALMA Observatory is operated by ESO, AUI/NRAO and NAOJ.
We thank the anonymous referee for the constructive comments, which helped strengthen this work.
\end{acknowledgement}

\bibliography{shocksulfur}
\bibliographystyle{aa} 

\begin{appendix} 

\section{Fitting an elliptical ring}\label{sec_fit_ring}

We modeled the ring-like emission as an elliptical Gaussian ring characterized by the intensity profile
\begin{equation}
I(x,y) =  H \exp\left[-\frac{1}{2}\left(\frac{r(x,y) - 1}{\sigma}\right)^2\right] + C,
\end{equation}
where $C$ is the average background intensity, $H$ is the peak intensity of the ring, and $\sigma$ represents the ring width normalized by the semi-major axis. The normalized elliptical radius $r(x,y)$ is defined as
\begin{equation}
r(x,y) = \sqrt{\left(\frac{x'}{a}\right)^2 + \left(\frac{y'}{b}\right)^2},
\end{equation}
where $a$ and $b$ are the semi-major and semi-minor axes of the ellipse, respectively. The rotated coordinates $(x', y')$ are obtained by rotating the pixel coordinates $(x,y)$ about the ring center $(x_0,y_0)$ by an azimuth angle $\phi_{\rm view}$:
\begin{equation}
\begin{pmatrix} x' \\ y' \end{pmatrix}
= 
\begin{pmatrix}
\cos \phi_{\rm view} & \sin \phi_{\rm view} \\
-\sin \phi_{\rm view} & \cos \phi_{\rm view}
\end{pmatrix}
\begin{pmatrix}
x - x_0 \\
y - y_0
\end{pmatrix}.
\end{equation}
The parameter set $\mathbf{p} = (x_0, y_0, a, b, \phi_{\rm view}, \sigma, H, C)$ was determined by fitting the model to the observed 2D emission map using a non-linear least squares optimization method. Initial guesses were based on visual inspection of the data.  
Fitting was performed using the \texttt{curve\_fit} routine from the \texttt{scipy} Python package, and uncertainties on the fitted parameters were estimated from the covariance matrix, with errors reported as one standard deviation (1$\sigma$).

\begin{figure}[!t]
    \centering
    \includegraphics[width=0.8\linewidth]{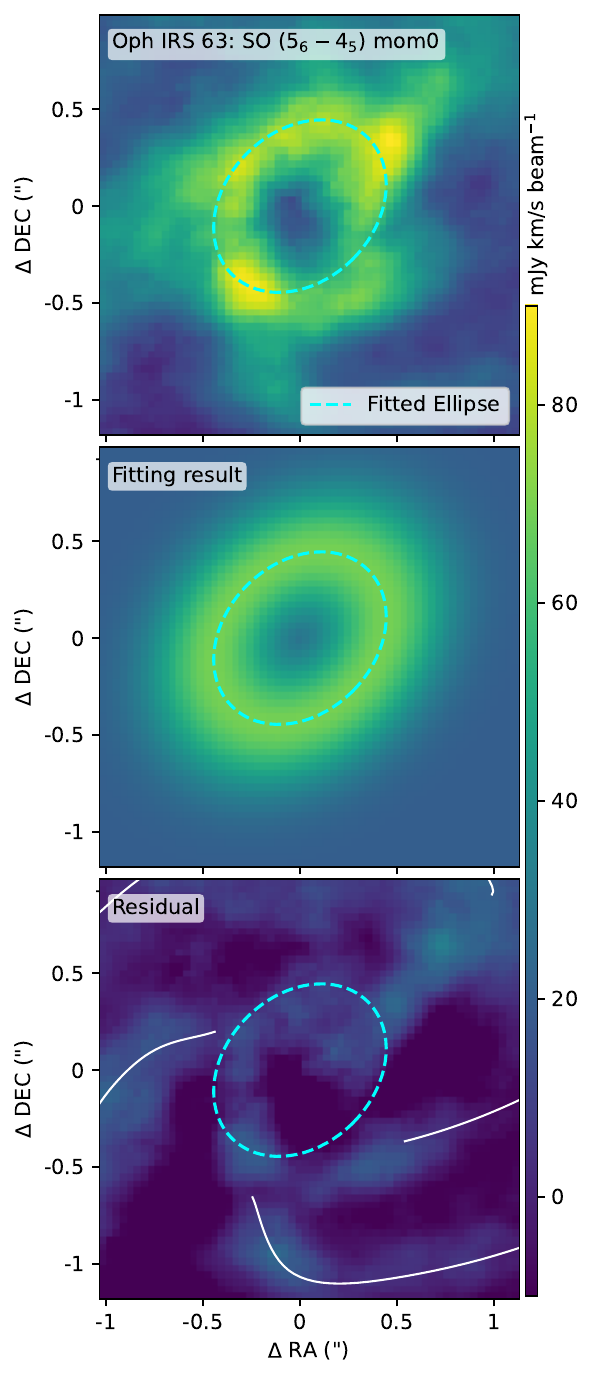}
    \caption{
Upper: Zoomed-in image of the SO ring-like structure of Oph IRS 63. 
Middle: The best-fit elliptical ring, as described in Sect.~\ref{sec_fit_ring}. 
Lower: Residuals of the SO emission after subtracting the fitted elliptical ring. 
The dashed cyan line marks the fitted ridge of the elliptical ring. The white lines indicate the four spiral-like structures (see Fig.~\ref{fig_irs63_color_maps}). Note that the field of view in these panels is smaller than in Fig.~\ref{fig_irs63_color_maps}.
    \label{fig:ringfit}}
\end{figure}

Applying this procedure to the moment 0 map of the SO emission in Oph IRS 63 yields the following best-fit parameters: semi-major axis $a = 0.50 \pm 0.01$ arcsec, semi-minor axis $b = 0.38 \pm 0.01$ arcsec, azimuth angle $\phi_{\rm view} = 52.93^\circ \pm 1.14^\circ$, normalized ring width $\sigma = 0.52 \pm 0.01$, peak intensity $H = 50 \pm 0.5$ mJy beam$^{-1}$ km s$^{-1}$, and background intensity $C = 20 \pm 0.5$ mJy beam$^{-1}$ km s$^{-1}$.  
The fitted model (Figure~\ref{fig:ringfit}) successfully reproduces the elliptical morphology of the ring emission, enabling quantitative characterization of its geometric and intensity structure.  
The inclination angle is estimated as $\theta_{\rm view} = \arccos(b/a) \approx 40.5^\circ \pm 1^\circ$.  
We note, however, that the presence of streamers and deviations from a perfectly elliptical ring may introduce systematic uncertainties not captured in the formal error estimates.  
The interaction between the spiral-like features—particularly S4—and the SO ring becomes more evident in the residual map after subtracting the fitted ring component (bottom panel of Fig.~\ref{fig:ringfit}).

\section{Extra images}\label{sec_ectraimage}
\begin{figure}[!t]
    \centering
    \includegraphics[width=0.76\linewidth]{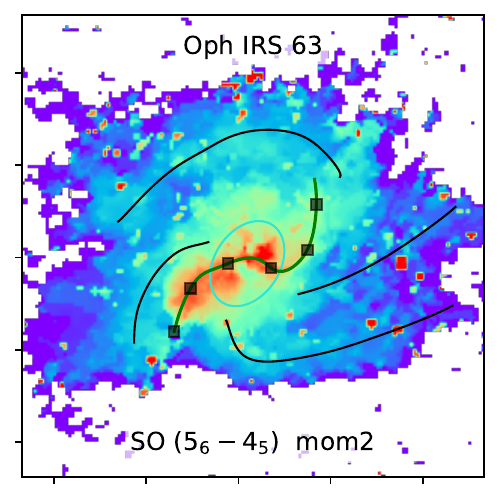}
    \caption{Moment 2 map of SO (5$_6$-4$_5$) toward Oph IRS 63, similar to panel (b) of Fig. \ref{fig_irs63_SO_mom12}. The green line along the black boxes marks the S-shaped feature observed in the moment 2 map. The black boxes indicate an angular separation of 0.5\arcsec, starting from the most bottom-left one.
    \label{fig_irs63_Smark}  }
\end{figure}

\begin{figure}[!t]
\centering
\includegraphics[width=0.9\linewidth]{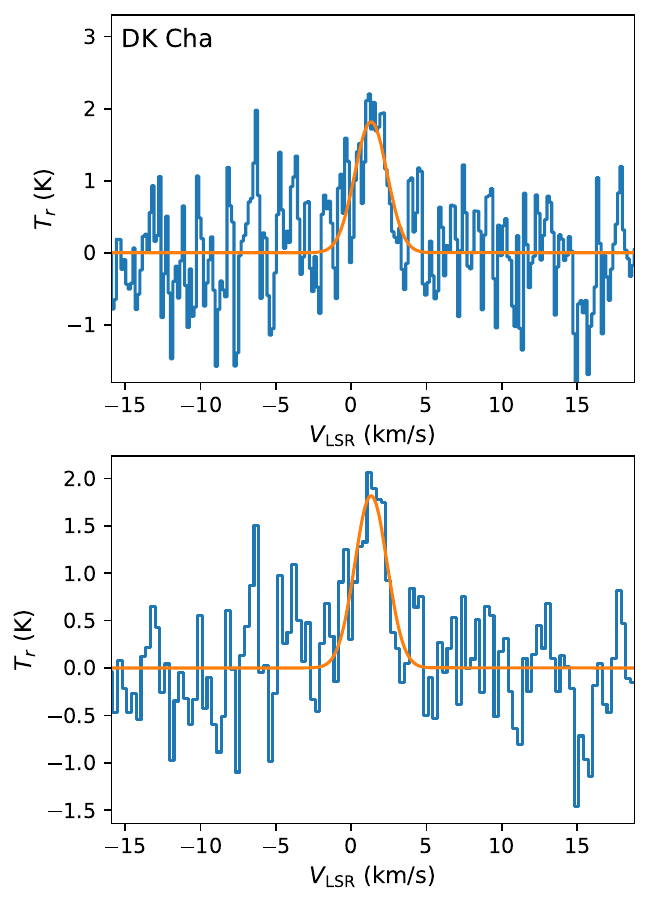}
\caption{Upper: The spectrum of SO$_2$ $16_{1,15}-15_{2,14}$ at position P2 of DK Cha (Fig. \ref{fig_moms_dkcha}d), supporting that P1 is a warm (and even hot) spot (Fig. \ref{fig_dkcha_warmspots}). The orange line shows the Gaussian fit. 
Lower: Same as the upper panel, but for the spectrum smoothed by convolving with a 3-point Hanning window.
\label{fig_dkcha_spw33spe} }
\end{figure}

\begin{figure}[!ht]
\centering
\includegraphics[width=0.98\linewidth]{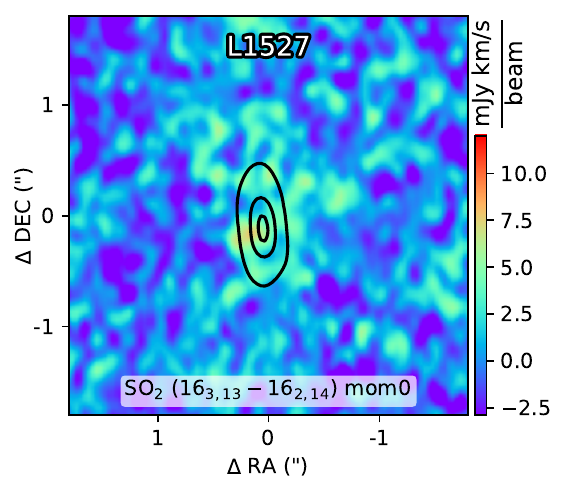}
\caption{Moment 0 map of SO$_2$ ($16_{3,13}-16_{2,14}$) of 
L1527. The black contours show the ALMA continuum at 1.3 mm 
same as the white contours in panels a--b of Fig. \ref{fig_L1527_mompv}.
\label{fig_L1527_SO2}}
\end{figure}

Images of observations referenced in this work but not included in the main figures will be presented here (Figs. \ref{fig_irs63_Smark},  \ref{fig_dkcha_spw33spe}, and \ref{fig_L1527_SO2}).

\section{Projection of parabolic trajectories} 
\label{sec_parabolic_trajectory}
\begin{figure}[!t]
    \centering
    \includegraphics[width=0.99\linewidth]{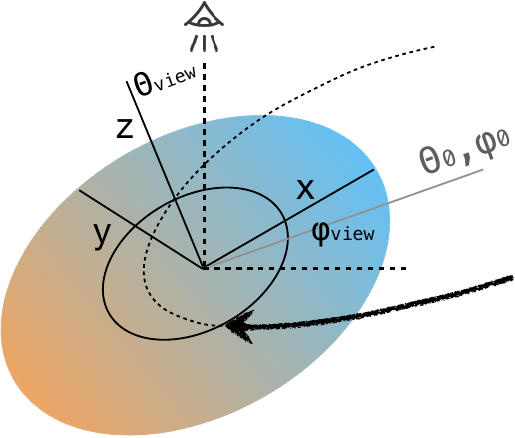}
    \caption{The schematic illustration of a free-fall stream with a parabolic trajectory (the black line with an arrow), initially originating from $(\theta_0, \phi_0)$ relative to the disk plane. The black ellipse marks the centrifugal barrier at a radius of $R_{\rm C}$ in the disk plane. The dashed curve represents the extended path of the parabolic trajectory, but the stream does not continue along the dashed curve as it is halted by the centrifugal barrier. Note that we adopt the right-handed coordinate system $(x, y, z)$, and in the case shown here, $\theta_{\rm view}$, $\phi_{\rm view}$, and $\phi_0$ all have negative values.
    \label{fig_stream_cartoon}
    }
\end{figure}

The parallel projection of a parabolic curve in 3D space onto a 2D plane still results in a parabolic curve. In the simplest case ($\theta_0 = \theta_{\rm view} = \phi_0 = \theta_{\rm view} = 0^\circ$, see Fig. \ref{fig_stream_cartoon}), the standard parabolic trajectory of a free-fall stream can be expressed as
\begin{equation}
    \frac{x}{R_{\rm C}} = \frac{1}{2} \left( \frac{y}{R_{\rm C}} \right)^2
\end{equation}
with $z=0$, $y < -R_{\rm C}$ in the clockwise case, and $y > R_{\rm C}$ in the counterclockwise case.
The free-fall velocity at $R_{\rm C}$ along the parabolic curve is
\begin{equation}
    v_{\rm C} = \left( \frac{2GM_\star}{R_{\rm C}} \right)^{1/2}.
\end{equation}
At any given point on the standard parabolic trajectory, the projected velocities along the $x$ and $y$ directions are given by
\begin{equation}
    v_x = \frac{y}{\sqrt{y^2 + R_{\rm C}^2}} \left( \frac{r}{R_{\rm C}} \right)^{-1/2} v_{\rm C}
\end{equation}
and
\begin{equation}
    v_y = \frac{R_{\rm C}}{\sqrt{y^2 + R_{\rm C}^2}} \left( \frac{r}{R_{\rm C}} \right)^{-1/2} v_{\rm C},
\end{equation}
respectively, with $r = \sqrt{x^2 + y^2}$, and $v_z = 0$.

Denote $\mathcal{M}^{j,k}$ as the value at the $j_{\rm th}$ row and $k_{\rm th}$ column of a matrix $\mathcal{M}$. Here, we define three matrices $\mathcal{R}_i$ for $i = 1, 2, 3$. For $i = 2$, the matrix is given by
\begin{equation}
\mathcal{R}_i(x) = \left|
\begin{array}{ccc}
\cos(x) & 0 & -\sin(x)  \\
0 & 1 & 0 \\
\sin(x) & 0 & \cos(x)
\end{array} \right|.
\end{equation}
That is, $R_i^{i,k} = R_i^{k,i} = 0$ for $k \neq i$, and $R_i^{i,i} = 1$. Similarly, $R_1$ and $R_3$ can be constructed.
Denote $\mathcal{A} = (x, y, z)$ as the points along the parabolic trajectory defined above. 
Let
\begin{equation}
\mathcal{R} = \mathcal{R}_3(\phi_{\rm view}) \cdot \mathcal{R}_1\left(\theta_{\rm view}\right) \cdot \mathcal{R}_3(\phi_0) \cdot \mathcal{R}_2\left(\frac{\pi}{2}-\theta_0\right).
\end{equation}
The transformation matrix $\mathcal{R}$ can generate the projection of any parabolic trajectory in 3D space through
\begin{equation}
\mathcal{A}_{\rm view}^{\rm T} = \mathcal{R} \cdot \mathcal{A}^{\rm T}.
\end{equation}
Similarly, the projection of the velocity along a parabolic trajectory, denoted as $\mathcal{V} = (v_x, v_y, v_z)$, can be expressed as
\begin{equation}
   \mathcal{V}_{\rm view}^{\rm T} = \mathcal{R} \cdot \mathcal{V}^{\rm T}.
\end{equation}
The observable parameters are the projected components $x$, $y$, and $v_z$.

For fixed values of $\theta_{\rm view}$ and $\phi_{\rm view}$, the free parameters of the trajectory ($\mathcal{A}_{\rm view}$) are $\theta_0$, $\phi_0$, and $R_{\rm C}$. The values of $\theta_{\rm view}$ and $\phi_{\rm view}$ can typically be determined from the morphology of the disk. Note that these values constrain the trajectory to intersect the midplane at a radius of $R_{\rm C}$. To obtain a parabolic curve without any constraints imposed by the morphology of the disk, the values of $\theta_{\rm view}$ and $\phi_{\rm view}$ may vary. Only four free parameters are needed to describe all parabolic curves with a fixed focus in 3D space: two parameters to define the plane of the curve, one to specify the axis of the parabola within the plane, and one to determine the focal length. By adopting different values for $(\theta_0, \phi_0, R_{\rm C}, \theta_{\rm view})$, we can generate all such parabolic curves. Fortunately, there are relatively simple methods for determining $\phi_{\rm view}$, such as measuring the large-scale velocity gradient or the direction of the outflows.

\section{Details of equations}\label{sec_eq_details}
For the CMU model \citep[][see also Sect. \ref{sec_cmu}]{1976ApJ...210..377U,1981Icar...48..353C}, to calculate the density $\rho$ (Eq. \ref{eq_rho_cmu}) at a given point \((r, \theta)\), the initial inclination angle of the streamline, $\theta_0$, should first be calculated from Eq. \ref{eq_sl}. This is a cubic equation in one variable. To solve Eq. \ref{eq_sl}, ChatGPT \citep{chatgpt2024} provides the solution
\begin{equation}
\mu_0 = \left\{ 
\begin{aligned}
&2\sqrt{\frac{-p}{3}}\cos\left(\arccos\left( \frac{3q}{2p\sqrt{-p/3}}
\right)/3\right)\ {\rm for}\ D<0, \\
&\left(-\frac{q}{2}+\sqrt{\frac{D}{108}}\right)^{\frac{1}{3}}+\left(-\frac{q}{2}-\sqrt{\frac{D}{108}}\right)^{\frac{1}{3}} \ {\rm for}\ D>0.
\end{aligned}
\right.  
\end{equation}
Here, $\mu=\cos(\theta)$, $\mu_0=\cos(\theta_0)$, $p=r/R_{\rm C}^{\rm mid}-1$, 
$q=-\mu r/R_{\rm C}^{\rm mid}$, and $D=4p^3+27q^2$.

For the constant-$J$ model (Sect. \ref{sec_hardsurface}), along a trajectory with $\theta_0 \neq 0\degr$, the azimuthal angle ($\phi$) can be expressed as a function of $r$ as
\begin{equation} \label{eq_phi}
\phi = \phi_0 \pm \arctan\left(  \frac{\sqrt{2r/R_{\rm C}-1}}{(r/R_{\rm C}-1)\sin(\theta_0)} \right).
\end{equation}
For any parabolic trajectory, when viewed from the pole, the increment of the azimuthal angle of the trajectory ($\Delta(\phi) = |\phi_0 - \phi|$) will always be $\leq 90^\circ$. However, when the line of sight is offset from the pole (even with a small $\theta_{\rm view}$), the increment of the projected azimuthal angle ($\Delta(\phi_{\rm proj})$) in some cases can be $\sim 180^\circ$ for a streamline with a small $\theta_0$. Therefore, in principle, $\Delta(\phi)$ can have an arbitrary value. Fortunately, if we constrain the trajectory to the midplane ($\theta_0 = 90^\circ$), we obtain the following constraint on $\Delta(\phi_{\rm proj})$:
\begin{equation}
\Delta(\phi_{\rm proj})  \leq 2\arctan\left( \frac{1}{\cos(\theta_{\rm view})}\right).
\label{eq_phi_incre}
\end{equation}
This can help quickly check whether a curved structure can be explained by a free-fall stream that flows in the midplane of the disk (Sect. \ref{sec_irs63_ring_steamer}).

In panel (b) of Fig. \ref{fig_irs63_pvmaps_ringandS}, the PV curve of a rotating ring is given by the following equations. The arc length of an ellipse is  
\begin{equation}
L(\phi) = a \cdot E\left(\phi + \frac{\pi}{2}, k\right) - a \cdot E\left(\frac{\pi}{2}, k\right). \label{eq_ellip_arclength}
\end{equation}
Here, $k = \sin(\theta_{\rm view})$, $a$ is the semi-major axis length, $E$ is the incomplete elliptic integral of the second kind, and $\phi$ is the azimuthal angle starting from one of the endpoints of the major axis. The projected velocity is expressed as 
\begin{equation}
V_{\rm proj} = V_{\rm rot} \cos(\phi) \sin(\theta_{\rm view}) + V_{\rm sys}. \label{eq_ellip_vproj}
\end{equation}
Here, $V_{\rm rot}$ is the rotational speed, and $V_{\rm sys}$ is the systemic velocity of the ring.

\section{Calculation of column density}\label{sec_cal_rtdiagram}
The column density of a species in the 
upper-level state ($N_u$) of a transition is \citep{2015PASP..127..266M}
\begin{equation}
    \frac{N_{\rm u}}{g_{\rm u}} = \frac{3k}{8\pi^3f_{\rm rest} S_{ij}\mu^2} \int T_{r} dV
\end{equation}
in the optically thin limit. Here, $k$ is the Boltzmann constant,
$T_{r}$ is the brightness temperature, $g_{\rm u}$ is the upper-level degeneracy,
and Debye$^2$ can be expressed in cgs unit as 
$10^{-36}\ \rm cm{^5} g^1\ s^{-2}$.
If $T_{r}$ is not small in comparison to the 
excitation temperature ($T_{\rm ex}$), 
a correction factor ($f_{\tau}$) should be applied, 
\begin{equation}
    f_{\tau} = \frac{\tau}{1-\exp(-\tau)}
\end{equation}
with 
\begin{equation}
    \tau = -\ln\left(1-\frac{T_r^{\rm peak}}{T_{\rm ex}} \right).
\end{equation}
Here, $T_r^{\rm peak}$ is the peak value  of the spectrum in 
brightness temperature. The total column density of the species
can be calculated through
\begin{equation}
    N = N_{\rm u} \frac{Q(T_{\rm ex})}{g_{\rm u}} \exp\left( \frac{E_{\rm u}}{T_{\rm ex}} \right). \label{Eq_N_total}
\end{equation}
Here, $Q(T_{\rm ex})$ is the partition function, which can be interpolated from the tabulated values available online \citep[e.g., through CDMS;][]{2001A&A...370L..49M}, and $E_{\rm u}$ is the upper-level energy.
Eq. \ref{Eq_N_total} yields \citep{1999ApJ...517..209G}
\begin{equation}
    \ln\left( \frac{N_{\rm u}}{g_{\rm u}} \right)=\ln\left( \frac{N}{Q(T_{\rm Tex})} \right) - \frac{E_{\rm u}}{T_{\rm ex}}. 
\end{equation}
If multiple transitions are available, the data in the $(E_{\rm u},\ \ln\left( \frac{N_{\rm u}}{g_{\rm u}} \right))$ space represent the rotational diagram. Linear fitting of the data yields the values of $N$ and $T_{\rm ex}$.

\section{Emission along outflow boundaries}\label{sect_lw_sw}
\begin{figure}[!t]
    \centering
    \includegraphics[width=0.95\linewidth]{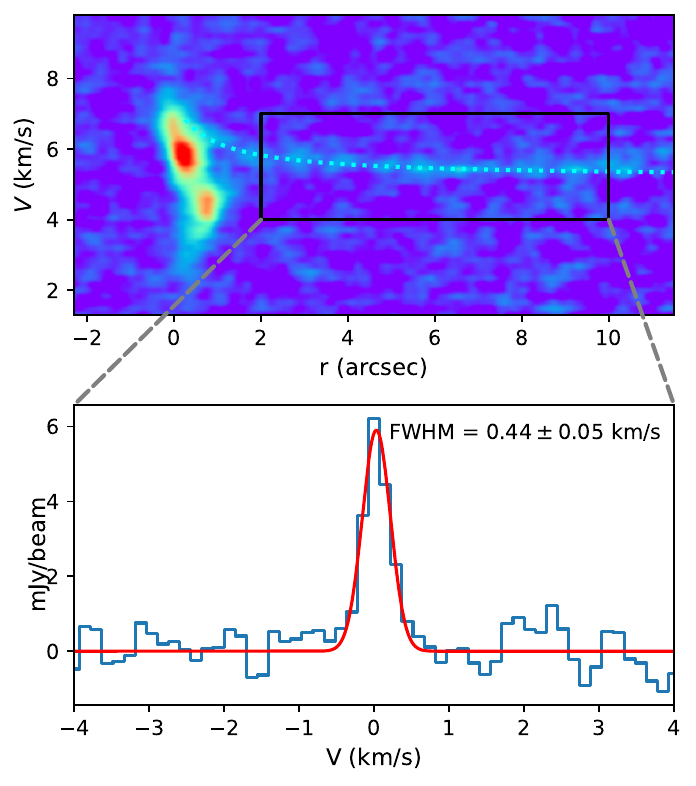}
    \caption{Upper: PV map of SO emission along the SW direction (same as Fig.~\ref{fig_L1527_cavity}(c)).  
Lower: Averaged spectrum (velocity centers of spectra at different $r$ have been corrected for better alignment; see Appendix~\ref{sec_eq_details} for details) of the region enclosed by the black rectangle in the upper panel. The red line shows the Gaussian fit, and the fitted linewidth (FWHM) is labeled.
}
    \label{fig:lw_sw}
\end{figure}

The outflow boundaries are manually drawn on the moment 0 map of CO (2--1) of L1527 (Fig.~\ref{fig_L1527_cavity}). To obatin smoothed boundary lines, cubic interpolation is then applied to the manually selected points along the boundaries (see the resultant black lines in Fig.~\ref{fig_L1527_cavity}).
\subsection{Narrow-line characteristics}
\begin{figure}[!t]
    \centering
    \includegraphics[width=0.96\linewidth]{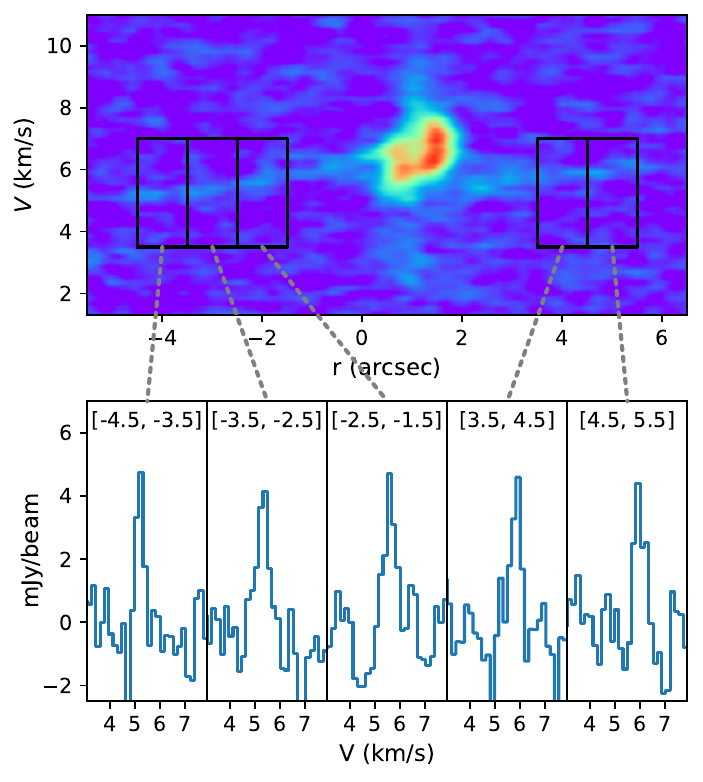}
    \caption{Upper: PV map of SO emission along the SE and NW directions (same as Fig.~\ref{fig_L1527_cavity}(d)).  
Lower: Stacked spectra averaged over different intervals of $r$, as indicated by the black rectangles in the upper panel. The corresponding intervals are also labeled on the individual spectra. Since the intervals are short, no velocity correction is applied here, unlike in Fig.~\ref{fig:lw_sw}.
}
    \label{fig:lw_se_nw}
\end{figure}

The upper panel of Fig.~\ref{fig:lw_sw} is identical to Fig.~\ref{fig_L1527_cavity}, showing the SO emission along the SW boundary of L1527 in the PV map. We denote \( V_0(r) \) as the fitted free-fall stream velocity in PV space. For each radius \( r \) within the main body of the SW region, enclosed by the black rectangle in the upper panel of Fig.~\ref{fig:lw_sw}, the spectrum along velocity \( V \) is resampled by subtracting \( V_0(r) \) from the velocity axis. This resampled spectrum is denoted as \( S_r \). 
The spectrum \( S_r \) is very narrow in line width (FWHM \( \sim 0.44 \) km s\(^{-1}\)), confirming the narrow-line feature of the SO strip and the good match with the modeled velocity distribution (dotted white line in Fig.~\ref{fig:lw_sw}). The stacked spectrum along the SW direction is obtained by combining all \( S_r \) over the radius \( r \). The lower panel of Fig.~\ref{fig:lw_sw} displays the stacked spectrum and its Gaussian fit, yielding a line width (FWHM) of \( 0.44 \pm 0.05 \) km s\(^{-1}\).
The SO emission along the SE–NW direction (Fig.~\ref{fig_L1527_cavity}) is also clearly visible in the PV map (Fig.~\ref{fig:lw_se_nw}), showing similarly narrow line widths.

\subsection{Emission along transverse direction}\label{sect_appendix_separation}
A line on a 2-D map, such as the outflow boundary, is denoted by a series of points \((x_i, y_i)\). The transverse direction at the \(i_{\rm th}\) point along this line is computed as
\begin{equation}
    \phi_{\rm trans} = \arctan\left( \frac{y_{i+1} - y_{i-1}}{x_{i+1} - x_{i-1}} \right) + \frac{n}{2}\pi,
\end{equation}
where \(n = 1\) or \(3\) is chosen to ensure that the transverse arrows point inward toward a specific direction, such as the outflow cavity (Fig.~\ref{fig_L1527_cavity}).

For the outflow boundary lines of L1527, we first calculate the transverse direction using the method described above. Emission profiles are then extracted along these transverse directions from the moment 0 maps. For both SO and CO, the resulting profiles—averaged over all transverse cuts—exhibit clearly spatially separated Gaussian components.
We denote the mean transverse offset of each Gaussian component as \(\mu\), with its associated uncertainty represented by \(\sigma_\mu\). To quantify the significance of the spatial separation between the CO and SO emission, we compute the \(t\)-statistic as
\begin{equation}
    t{\rm -stat} = \frac{\left| \mu^{\rm CO} - \mu^{\rm SO} \right|}{ \sqrt{  \sigma^{2}_{\mu^{\rm CO}} + \sigma^{2}_{\mu^{\rm SO}} } }.
\end{equation}
Table~\ref{tab_L1527_offset} lists the measured transverse separations between the SO and CO emission for the four outflow boundaries in L1527, as illustrated in Fig.~\ref{fig_so_co_offset}.

\begin{table}[!h]
\caption{Separation between the SO and CO emission along the 
transverse directions of the outflow boundaries in L1527 (Fig.~\ref{fig_so_co_offset}).
\label{tab_L1527_offset}}
\centering
\begin{tabular}{lcccc}
\hline\hline
Region & $\mu^{\mathrm{SO}}$ ($\arcsec$) & $\mu^{\mathrm{CO}}$ ($\arcsec$) & $\mu^{\mathrm{CO}} - \mu^{\mathrm{SO}}$ ($\arcsec$) & $t$-stat \\
\hline
NE & $-0.18 \pm 0.03$ & $0.60 \pm 0.04$ & $0.78 \pm 0.05$ & 15.3 \\
NW & $-0.09 \pm 0.06$ & $0.75 \pm 0.03$ & $0.84 \pm 0.06$ & 14.2 \\
SE & $0.14 \pm 0.13$  & $0.63 \pm 0.03$ & $0.48 \pm 0.13$ & 3.7  \\
SW & $-0.46 \pm 0.03$ & $0.30 \pm 0.04$ & $0.75 \pm 0.05$ & 15.0 \\
\hline
\end{tabular}
\vspace{0.5em}
\begin{minipage}{0.9\linewidth}
\footnotesize
\textit{Note.} See Appendix~\ref{sect_appendix_separation} for the definitions of $\mu$ and the $t$-statistic.
\end{minipage}
\end{table}

\end{appendix}
%
%
\end{document}